\documentclass[useAMS,usenatbib]{mn2e}
\usepackage{amsmath}
\usepackage{amssymb}
\usepackage{multirow}
\usepackage{graphicx}
\usepackage{epstopdf}

\usepackage{color,url}

\def\simlt{\lower.5ex\hbox{$\; \buildrel < \over \sim \;$}}
\def\simgt{\lower.5ex\hbox{$\; \buildrel > \over \sim \;$}}

\newcommand{\bd}{\begin{displaymath}}
\newcommand{\ed}{\end{displaymath}}
\newcommand{\be}{\begin{equation}}
\newcommand{\ee}{\end{equation}}
\newcommand{\beqa}{\begin{eqnarray}}
\newcommand{\eeqa}{\end{eqnarray}}

\title[Global Signal from 21-cm Fluctuations] 
{Extracting the Global Signal from 21-cm Fluctuations: the Multi-Tracer Approach} 
\author[Fialkov, Barkana \& Jarvis] {Anastasia Fialkov$^{1,2}$\thanks{E-mail:
    a.fialkov@sussex.ac.uk},
 Rennan Barkana$^{3,4}$, Matt Jarvis$^{4,5}$ \\
$^{1}$Institute of Astronomy, University of Cambridge, Madingley Road, Cambridge CB3 0HA, UK\\
$^{2}$Department of Physics and Astronomy, University of Sussex, Falmer,  Brighton BN1 9QH, UK\\
  $^{3}$ Raymond and Beverly Sackler School of Physics and Astronomy,
    Tel Aviv University, Tel Aviv 69978, Israel\\
        $^{4}$ Department of
        Astrophysics, University of Oxford, Denys Wilkinson Building,
        Keble Road, Oxford OX1 3RH, UK\\
        $^{5}$ Department of Physics \& Astronomy, University of the Western Cape, Private Bag X17, Bellville, Cape Town, 7535, South Africa}

\begin{document}
\pagerange{\pageref{firstpage}--\pageref{lastpage}} \pubyear{2019}
\maketitle

\label{firstpage}

\begin{abstract}
The multi-tracer technique employs a ratio of densities of two differently biased galaxy samples
that trace the same underlying matter density field,
and was proposed to alleviate the cosmic variance problem. Here we
propose a novel application of this approach, applying it to two
different tracers one of which is the 21-cm signal of neutral hydrogen
from the epochs of reionization and comic dawn. The second tracer is
assumed to be a sample of high-redshift galaxies, but the approach can
be generalized and applied to other high-redshift tracers. We show
that the anisotropy of the ratio of the two density fields can be used
to measure the sky-averaged 21-cm signal, probe the spectral energy
distribution of radiative sources that drive this signal, and extract
large-scale properties of the second tracer, e.g., the galaxy
bias. Using simulated 21-cm maps and mock galaxy samples, we find that
the method works well for an idealized galaxy survey. However, in the case of
a  more realistic galaxy survey which only probes highly biased luminous galaxies, 
the inevitable Poisson noise makes  the reconstruction  far more challenging. 
 This difficulty can be mitigated with the greater sensitivity of future telescopes along
with larger survey volumes.

\end{abstract}


\section{Introduction}
\label{Sec:Intro}

The exploration of the first billion years of cosmic history is under way: observations of lensed fields with the  Hubble Space Telescope (HST) have revealed bright  galaxies at the onset of the Epoch of Reionization (EoR, $z\sim 6-10$)  with the most distant galaxy detected at redshift $z\sim 11.1$ \citep{Oesch:2016}; the Atacama Large Millimeter Array (ALMA) sees dusty bright galaxies at $z\sim 9$ \citep[e.g.,][]{Hashimoto:2018}; while high-redshift  quasars have been seen above redshift  $7$ \citep[the highest redshift quasar was found at $z = 7.54$,][]{Banados:2018}. Existence of such massive and metal-rich objects when the Universe was less than a billion years old suggests an early formation of the first stars \citep[e.g., as early as $z\sim 15$ in the case of the metal-rich galaxy at $z=9.1$,][]{Hashimoto:2018}. 

According to theory, the first stars turn on at $z\sim 30-60$ \citep[e.g.,][]{Naoz:2006, Bromm:2017}. This hypothesis can be tested with the next generation instruments designed to observe the high-redshift Universe at different wavelengths. Some examples include the effort in X-rays such as {\it Lynx} \citep{Gaskin:2018}, which is predicted to have enough sensitivity to see accreting black holes of mass $10^4$ M$_\odot$ at $z\sim 10$, and infrared  telescopes such as the {\it James Webb Space Telescope} \citep[{\it JWST}, which should have enough sensitivity to see galaxies out to redshift $\sim 16$,][]{Cowley:2018}.  However, these telescopes will have small fields of view and will mainly yield deep observations of small patches of the sky.  Intensity mapping of molecular lines is another promising technique to probe the  typical population of high-redshift star forming galaxies, allowing us
to survey large cosmic volumes \citep[e.g.,][]{Kovetz:2018, Moradinezhad:2019b, Moradinezhad:2019}. 

The 21-cm signal produced by neutral hydrogen atoms in the Intergalactic Medium (IGM) is predicted to be a powerful probe of the early Universe at $z\gtrsim 6$ and is  complementary to the direct observations of bright sources \citep[see][for a recent review of the 21-cm line]{Barkana:2016}. The brightness temperature of the 21-cm line is driven by the thermal and ionization histories of the gas as well as by the Ly-$\alpha$ radiation of the first  stars \citep[Wouthuysen-Field effect,][]{Wouthuysen:1952, Field:1958}.  
 
The sky-averaged (global) 21-cm signal   contains information on the evolution history of the Universe. The timing of cosmic events such as the formation of the  first population of stars, the onset of cosmic heating by the first population of X-ray sources, and the beginning of reionization are imprinted in the shape of the global signal \citep[e.g.,][]{Cohen:2017}. Owing to the high expected scientific gains, several pioneering instruments are aspiring to observe  the all-sky radio spectrum \citep{Bowman:2010, Philip:2018, Price:2018, Singh:2017a, Voytek:2014}.  Recently, the EDGES team  has claimed discovery of the cosmological signal from  redshifts  $z\sim 14-26$. An absorption trough with amplitude close to 500 mK  (three times deeper than expected in standard astrophysical scenarios) centered at 78 MHz was measured with EDGES Low-Band instruments \citep{Bowman:2018}.   If confirmed\footnote{Concerns were raised by  \citet{Hills:2018}, which the EDGES team has addressed \citep{Bowman:2018}. Reanalysis of the integrated publicly available spectrum was also done by \citet{Singh2019}.}, this signal is the first observational evidence of primordial star formation at the dawn of the Universe when it was only 200 million years old.   At lower redshifts, attempts to observe the global signal have so far yielded non-detection, placing limits on the astrophysical parameter space  \citep{Monsalve:2017b, Monsalve:2018a, Monsalve:2018b, Singh:2017a, Singh:2017b}. 

Fluctuations in the 21-cm field  are  complementary to the global signal, containing far more information on the distribution of radiative sources as well as on the characteristic scales on which these  sources affect the environment. The latter can be related to the typical wavelength of the emitted photons, making the fluctuations a sensitive probe of the nature of the first stars and black holes \citep[e.g., ][]{Fialkov:2014, Fialkov:2014b}. The first attempt to detect fluctuations in the 21-cm signal via the redshifted 21-cm line at 151 MHz  dates back to 1986 and targeted gaseous pancakes at $z\sim8$  \citep{Bebbington:1986}. At present, interferometric arrays are targeting fluctuations of the signal from the EoR as well as from the earlier epoch of cosmic dawn, yielding upper limits \citep[e.g.,][]{Paciga:2013, Beardsley:2016, Patil:2017,  Gehlot:2018, Eastwood:2019}. Although at present these limits are too weak to have interesting astrophysical implications, the effort is ongoing. Dedicated experiments such as the Hydrogen Epoch of Reionization Array \citep[HERA,][]{DeBoer:2017} and facilities such as  NenuFAR \citep{Zarka:2012} and the Square Kilometer Array  \citep[SKA,][]{Koopmans:2015} will measure the 21-cm fluctuations from cosmic dawn and the EoR over a wide range of scales. %
 
Expected correlations between the features of the global signal and the details of the power spectrum \citep{Cohen:2018} suggest that the power spectrum measurements could be used to infer the global history of the Universe. Extracting the global signal directly  from the interferometric measurements of the power spectrum has been discussed in the literature \citep{Presley:2015, Singh:2015, Venumadhav:2016, McKinley:2018}. However, such  observations  are particularly challenging because, in order to extract the global signal from the visibilities, one needs to characterize the antenna cross-talk and the correlated thermal noise to high precision \citep{Bernardi:2015, Presley:2015, Singh:2015, Venumadhav:2016}. In this paper we propose an alternative method to measure the global signal from the 21-cm fluctuations using the multi-tracer approach.  The paper is organized as follows: in Section \ref{Sec:Method} we briefly summarize the standard multi-tracer technique \citep{Seljak:2009}. The  physics of the 21-cm line, as well as the simulations used to generate the 21-cm signals and the galaxy fields, are described in Section \ref{Sec:Signals}. We summarize the parameter extraction algorithm  in Section \ref{Sec:Fitting}. The results are reported in Section \ref{Sec:Res}.  Finally, we conclude in Section \ref{Sec:sum}.   Throughout the paper we assume standard $\Lambda$CDM cosmology with the values of cosmological parameters measured by the {\it Planck} satellite \citep{Planck:2016}. All distances are calculated in comoving Mpc and wavenumbers in comoving Mpc$^{-1}$. Because in this paper  we are mainly interested in illustrating a new technique and are predominantly focusing on low redshifts ($6\lesssim z\lesssim 15$), we do not address the EDGES-Low observation and only  consider  standard astrophysical models \citep[e.g., as is summarized in][]{Cohen:2017}.  However, if the EDGES detection is confirmed, the unusually strong global 21-cm signal would imply  enhanced fluctuations \citep[e.g.,][]{Fialkov:2018, Fialkov:2019} and make the proposed  measurement much more feasible.

\section{Multi-Tracer Technique}
\label{Sec:Method} 

Originally introduced  to compensate for the cosmic variance in  large scale structure galaxy surveys \citep{Seljak:2009}, the multi-tracer method employs  a ratio of the density fields of two different  populations of galaxies, $\delta_{\rm gal}({\bf k},z)$,  tracing the same underlying large-scale matter density field, $\delta({\bf k},z)$. In the linear regime the relationship between the galaxy and matter density fields in Fourier space is 
\begin{equation}
\delta_{\rm gal}({\bf k},z)=\left(b+\mu^2_{{\bf k}} f\right) \delta({\bf k},z),
\label{dgal}
\end{equation}
where ${\bf k}$ is the comoving wavenumber,  $b$ is the bias of the galaxy field, $\mu_{\bf {k}} = \cos \theta_{\bf {k}}$ is the cosine of the angle $\theta_{\bf k}$ with respect to the  line of sight, and the factor $f = d\ln D/d\ln (1+z)^{-1}$ measures the growth of structure and is close to unity at the redshifts of interest ($6\lesssim z\lesssim 30$) where the Universe is matter dominated.  In Eq.   \ref{dgal} the angular dependence is introduced by the Kaiser effect  \citep{Kaiser:1987}. Because the two galaxy  populations  sample the  same underlying density field,  the ratio of the Fourier transforms of their densities is independent of $\delta({\bf k},z)$, and for every $\mu_{\bf {k}}$ it is fully determined by the galaxy biases (and the factor $f$).  Therefore, the ratio does not suffer from cosmic variance and allows access to information on the largest observable scales where the number of samples is small. This method has been shown to improve the constraints on non-Gaussianity of the initial conditions from inflation \citep{Seljak:2009},  redshift space distortions \citep{McDonald:2009}, general-relativistic effects in the observed density of sources \citep{Alonso:2015}.

In this paper we develop a new application of the multi-tracer method by  replacing one of the tracers by simulated fluctuations in the 21-cm signal, $\delta_{21}({\bf k},z)$.  The resulting ratio of the density fields  is deterministic and anisotropic. We show that the  angular dependence  of the ratio  can be used to extract the global 21-cm signal, galaxy bias and spectral properties of the first X-ray sources.
 
\section{Simulated Data}
\label{Sec:Signals}

\subsection{21-cm Signal}

The temporal evolution of the real-space three dimensional differential brightness temperature of the 21-cm signal  seen against the Cosmic Microwave Background (CMB) radiation at low radio frequencies is given  by
\begin{equation}
T_{21}({\bf x},z)  = \frac{T_S-T_{\rm CMB}}{1+z}\left[1-\exp^{-\tau_{21}}\right],
\label{Eq:T}  
\end{equation}
where ${\bf x}$ is the comoving coordinate,  $T_S$  is the spin temperature of the 21-cm transition \citep[e.g.,][ and references therein]{Barkana:2016},  $T_{\rm CMB} = 2.725(1+z)$ K is the CMB temperature at redshift $z$, and $\tau_{21}$ is the 21-cm optical depth given by
\begin{equation}
\tau_{21}  = \frac{3h_{\rm pl}A_{10}c \lambda_{21}^2n_{\rm HI}}{32\pi k_B T_S (1+z) dv_r/dr}.
\label{Eq:tau}
\end{equation}
Here $h_{\rm pl}$ is Planck's constant,  $A_{10}=2.85\times 10^{-15}$ s$^{-1}$ is the spontaneous emission coefficient, $c$ is the speed of light, $\lambda_{21} =21$ cm is the rest-frame wavelength of the signal,  $n_{\rm HI}$ is the number density of neutral hydrogen atoms which depends on the cosmological parameters  and ionization history, $k_B$ is the Boltzmann constant, and $dv_r/dr$ is the  radial component of the velocity gradient created by structure formation which (on average) equals   $dv_r/dr = H(z)(1+z)^{-1}$ where $H(z)$ is the Hubble parameter. Throughout this paper we assume  that the 21-cm optical depth  is small.

The spatially averaged $T_{21}({\bf x},z)$ gives the global 21-cm signal,  $T_{21}(z)$. The density contrast in Fourier space,  $\delta_{21}({\bf k},z)$, in mK units is the Fourier transform of $T_{21}({\bf x},z)-T_{21}(z)$. 

\subsubsection{Cosmic History}
\label{Sec:history}

The dependence of $T_{21}({\bf x},z)$  on astrophysical parameters is encoded via their effects on the spin temperature  and the ionization history. The variation of the 21-cm signal as a function of astrophysical parameters was recently explored  by  \citet{Cohen:2017, Cohen:2018, Cohen:2019} who studied a large sample of 21-cm signals,  with the key astrophysical parameters varied over the widest range allowed by existing observational and theoretical constraints.

In particular, Ly$-\alpha$ radiation produced by the first stars couples the spin temperature to the kinetic temperature of the gas, $T_K$ \citep{Wouthuysen:1952, Field:1958}. The exact timing of the Ly$-\alpha$ coupling depends on the dominant regime of star formation (e.g., atomic cooling takes place in dark matter halos above the mass threshold of  $\sim 10^7$ M$_\odot$) as well as on the star formation efficiency, $f_*$. This process is very uncertain due to the lack of observations of high-redshift stars and is predicted to  occur between  $z\sim 35$ and $z \sim 12$  \citep{Cohen:2018}. At these early times, after the onset of star formation and prior to the build-up of  the first population of X-ray sources, fluctuations in the 21-cm signal are predominantly driven by the nonuniform Ly$-\alpha$ background \citep{Barkana:2005b, Cohen:2018}. The peak power of these fluctuations at $k = 0.1$ Mpc$^{-1}$ falls in the range $z\sim 14-35$ \citep{Cohen:2018}. As more and more stars form, the intensity of the Ly$-\alpha$ flux rises and its effect on the 21-cm signal saturates, leading to a suppression of the power spectrum. Although in this paper we focus on the effect of X-ray sources and ionizing radiation (i.e., lower redshifts), the process of Ly$-\alpha$ coupling is modeled self-consistently and our findings can be equally applied to the epoch of  Ly$-\alpha$ coupling.

At the dawn of star formation intergalactic neutral gas is much colder than the CMB as a result of the cosmic expansion and due to the lack of significant X-ray heating. Therefore, the 21-cm signal is expected to be seen in absorption against the background. As soon as the first X-ray sources emerge, they heat up the gas in a non-uniform way.  The  heating transition (occurring at redshift  $z_h$) marks the moment when the 21-cm signal vanishes as a result of the average gas temperature nearing that of  the background radiation. At redshifts lower than $z_h$, the 21-cm signal is observed in emission. According to the parameter study by \citet{Cohen:2018},   $z_h$ is lower than 22 for all the explored scenarios. In some models,  X-ray heating is never strong enough to heat the gas above $T_{\rm CMB}$.  Non-uniform heating leads to a boost in power of the 21-cm fluctuations \citep{Pritchard:2007} and is the primary source of fluctuations at $z\sim 9-24$ \citep{Cohen:2018}. Properties of X-ray sources, such as their bolometric luminosity and spectral energy distribution (SED), affect the heating process, and, therefore, can be extracted from the 21-cm signal. In particular, in \citet{Fialkov:2014} and \citet{Fialkov:2014b} we compared the effect of a realistic SED of X-ray binaries  \citep[a relatively hard spectrum peaking at $\sim 2$ keV, ][]{Fragos:2013} to the heating by softer sources. We found that hard sources are less efficient in heating up the gas, which  results in a  deeper absorption trough, lower emission peak and lower $z_h$. The SED has a strong effect on fluctuations: the power is suppressed on scales smaller than the mean free path of X-ray photons. In the case of the soft SED the mean free path is short and most of the injected energy is absorbed close to the sources, while in the case of the hard sources heating is distributed over a larger range of scales.

The effect of X-rays on the 21-cm signal saturates once the gas temperature exceeds that of the CMB, and at later times the 21-cm signal is driven mostly by the reionization history.
The process of reionization dominates  the low-redshift regime, with the fluctuations peaking at $z<13$ \citep[again, the exact timing is very model dependent, ][]{Cohen:2018}. According to the existing observations of quasars, Ly$-\alpha$ emitters, Lyman-break galaxies and the CMB optical depth, reionization ends at $z_{\rm eor}\sim 6$ or later \citep{McGreer:2015, Greig:2017, Banados:2018, Mason:2018, Planck:2018, Weinberger:2019} and the 21-cm signal from the IGM essentially vanishes at that redshift.

\subsubsection{Simulations}
\label{sec:Mdata}
Three-dimensional realizations of the evolving 21-cm signal are generated using hybrid  simulations  \citep[e.g.,][]{Fialkov:2014,Cohen:2017} inspired by \citet{Mesinger:2011}. The simulations were initiated at $z = 60$, at which the real-space density field $\delta({\bf x},z)$ was seeded  with the resolution of 3 comoving Mpc, and evolved down to $z=6$. The density fluctuations on the 3 Mpc scale are evolved linearly with redshift,  while sub-grid models are used to account for processes occurring at unresolved scales, such as gravitational collapse and star formation. Abundance of dark matter halos at  each redshift and of each halo
mass M$_{\rm h}$ is computed using the extended Press-Shechter
formalism \citep{Barkana:2004}.  Each halo above the atomic cooling threshold is assumed to host a galaxy of stellar mass M$_* = f_*f_{\rm gas}$M$_{\rm h}$   where $f_{\rm gas}$ is the redshift-dependent fraction of the baryon density contained in haloes of mass M$_{\rm h}$ which also depends on the local value of the relative velocity between dark matter and gas \citep{Tseliakhovich:2011}.   We assume  Population II star formation with star formation efficiency of $f_*= 5\%$. \citep[See, e.g., ][for alternative galaxy models used in other 21-cm studies.]{Mirocha:2017, Mirocha:2018} Radiative backgrounds produced by stars and their remnants are calculated accounting for inhomogeneity and light-cone effects. Reionization is modelled using the excursion set formalism \citep{Furlanetto:2004}.  Finally, using the calculated inhomogeneous time-dependent fields (density, X-ray and  Ly$-\alpha$ backgrounds) as well as reionization history, the evolution of the redshifted isotropic 21-cm signal is followed according to Eq. \ref{Eq:T} in the limit of small optical depth. 
Peculiar velocity effects are added in the post-processing stage as described in Section \ref{Sec:anis}.  

Our results are shown for three combinations of reionization and heating parameters \citep[identical to what we explored in ][]{Fialkov:2015}:
\begin{itemize}
\item  Case1:  Late EoR ($\tau = 0.067$, $z_{\rm eor}\sim 6.5$) and soft X-ray SED. The redshift of the heating transition in this case is $z_h = 14.5$.
\item  Case2: Late EoR ($\tau = 0.067$, $z_{\rm eor}\sim 6.5$) and hard X-ray SED with $z_h = 12.1$.
\item  Case3: Early EoR ($\tau = 0.085$, $z_{\rm eor}\sim 8$) and soft X-rays yielding $z_h = 14.5$.
\end{itemize}
All cases assume an identical X-ray bolometric luminosity per star formation  rate of $3\times10^{40}$ erg s$^{-1}$ M$_{\odot}^{-1}$ yr. 

\begin{figure}
\includegraphics[width=3.4in]{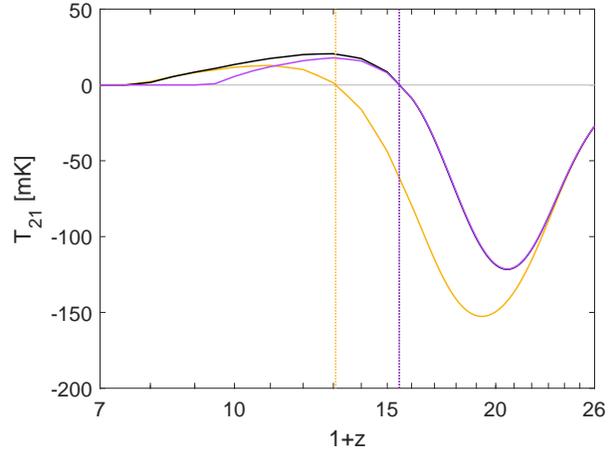}
\caption{Simulated global 21-cm signals for  Case1 (black), Case2   (orange),   Case3 (purple). Vertical dotted lines show the redshifts of the heating transition. Horizontal dotted line shows marks $T_{21}=0$ for reference.}
\label{Fig:T1}
\end{figure}

The corresponding 21-cm global signals are shown in Figure \ref{Fig:T1} (solid lines). As expected, at high redshifts the signals  1 and 3 coincide  because of the identical X-ray heating, differing only at lower redshifts when reionization becomes important. On the contrary, signals 1 and 2 are identical at low reshifts  and differ at high-$z$. This is because  Case1 and Case2 have identical reionization histories but different X-ray heating, and at low-$z$  X-ray heating is saturated and does not play a role.  The absorption trough is much deeper in Case2 compared to Cases 1 \& 3 because hard X-ray sources are less efficient in heating the IGM than the soft sources \citep[e.g.,][]{Fialkov:2014}.

Using the specifications of future facilities (such as the SKA\footnote{SKA will have a large field of view of $\sim 5$ degrees at 100 MHz, corresponding to $\sim 823$ comoving Mpc at $z = 9$}) as an indication,  we employ cosmological  simulations of  768  Mpc on a side  with  3 Mpc resolution, which give access to fluctuations on scales $\sim 0.01-1$ Mpc$^{-1}$. We also use smaller simulations of 384$^3$ Mpc$^3$ to explore scaling of the statistical errors with the survey volume  (see discussion in Section \ref{Sec:Fitting}). 
 
\subsubsection{Anisotropy}
\label{Sec:anis}
To the leading order, the 21-cm signal (Eq. \ref{Eq:T}) is isotropic and  depends on the evolving distribution of star forming halos  convolved with the spherically symmetric window functions, $W(x,z)$, which quantify the response of the signal to the radiative backgrounds \citep[X-ray, UV and Ly-$\alpha$,][]{Barkana:2005, Fialkov:2015}.  In Fourier space and in the linear regime the signal can be written as 
\begin{equation}
\delta_{21}^{\rm iso}({\bf k},z) =  T_{21}W(k,z)\delta({\bf k},z),
\label{Eq:iso}
\end{equation}
where $W(k,z)$ is the Fourier transform of $W(x,z)$. We calculate the window function directly from the simulated data (prior to adding the velocity effects) by averaging the ratio  $T_{21}^{-1}\delta_{21}({\bf k},z)/\delta({\bf k},z)$  over the direction of ${\bf k}$ and refer to it as the ``theoretical'' window function.

\begin{figure*}
\includegraphics[width=3.4in]{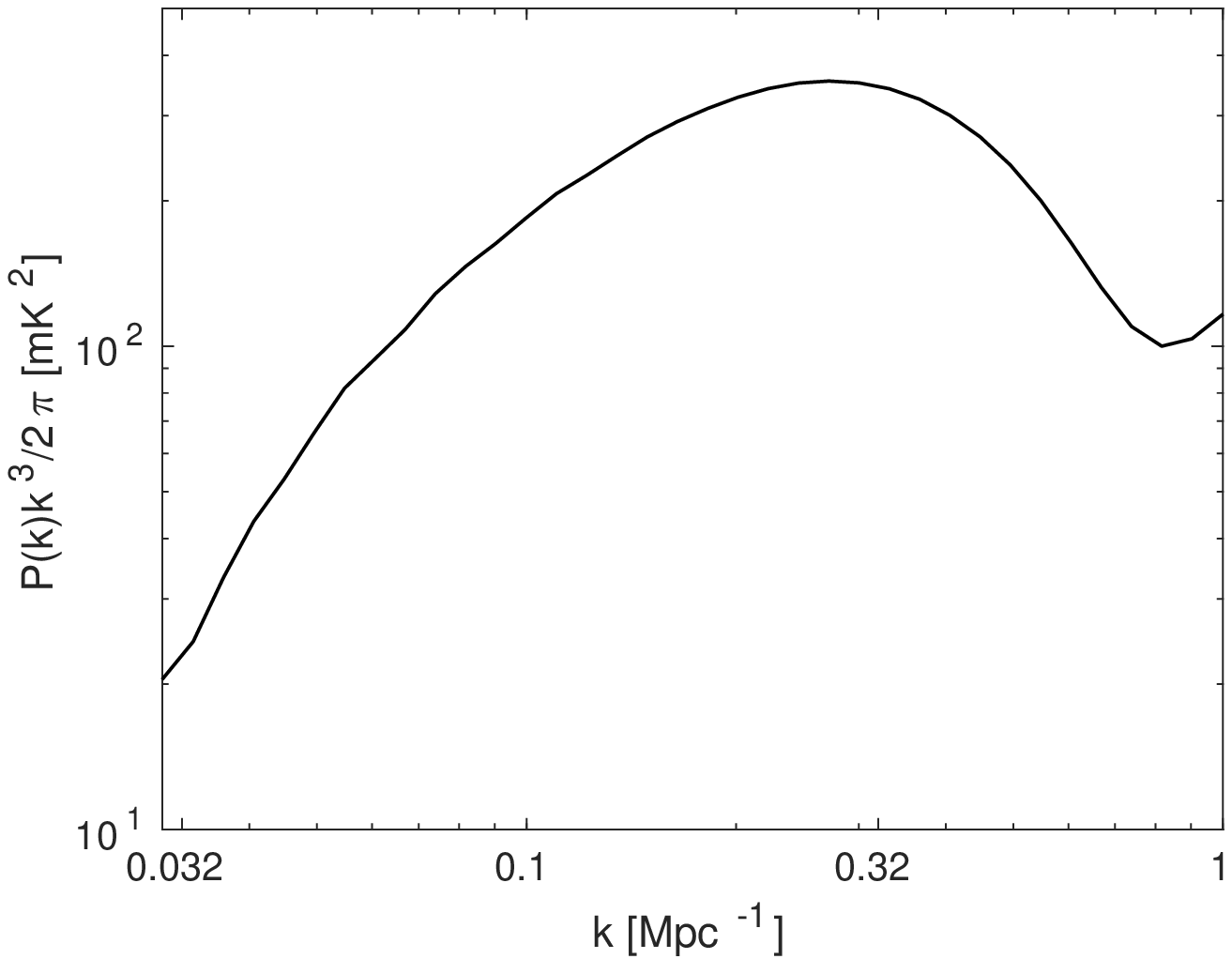}\includegraphics[width=3.4in]{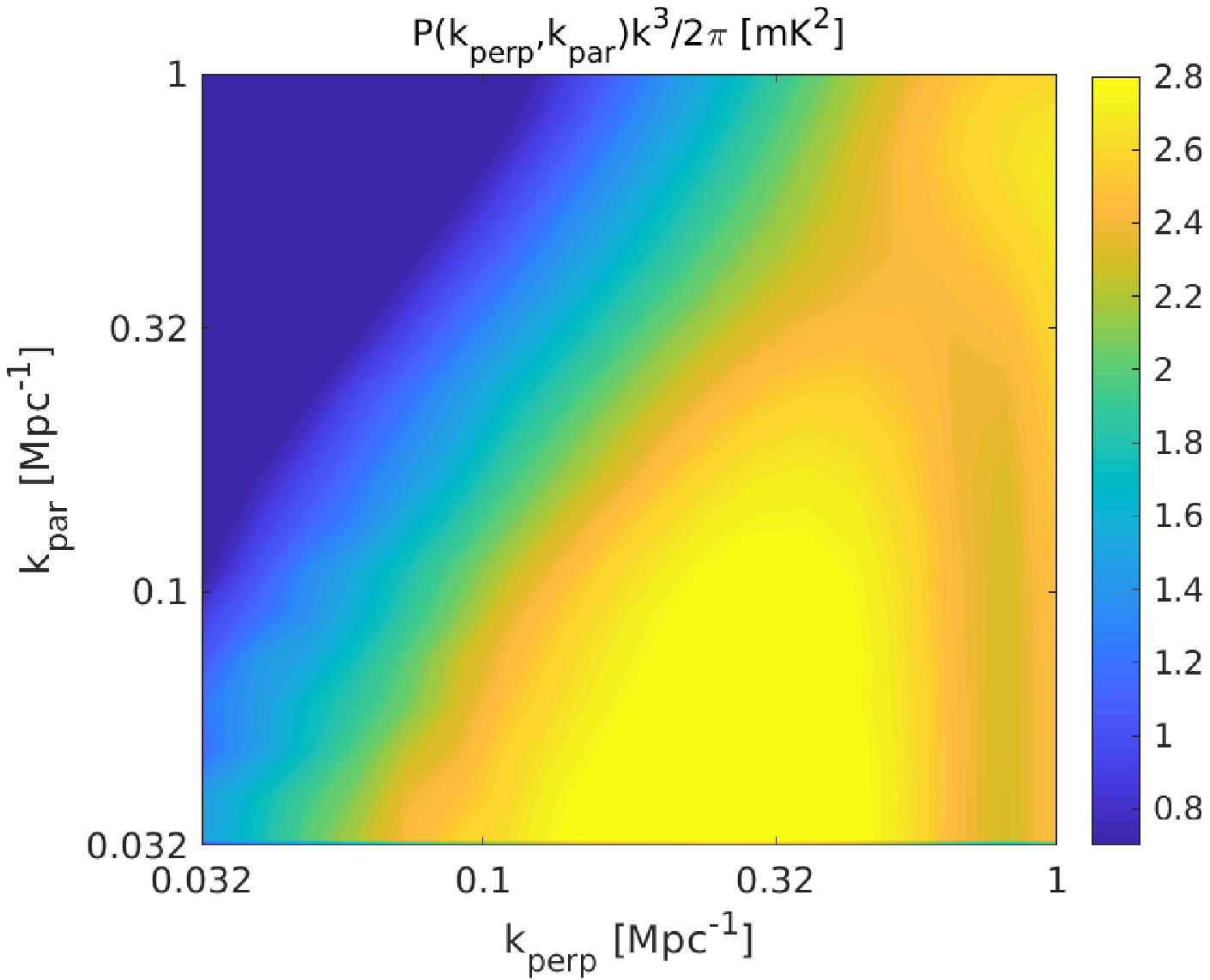}
\caption{Power spectra of 21-cm fluctuations for Case1 at $z=19$. We show spherically averaged isotropic power spectrum versus $k$ (left) and two-dimensional power spectrum as a function of $k_{\rm perp} = k\sqrt{1-\mu^2}$  and the absolute value of $k_{\rm par} = \mu k$  (right). Color code (see the colorbar on the right) corresponds to the logarithm (base 10) of the anisotropic power spectrum $P(k_{\rm perp},k_{\rm par})k^3/2\pi$ in mK units. The velocity field was added non-linearly.}
\label{fig:PS}
\end{figure*}

The peculiar velocity field  adds non-linearity, breaks the spherical symmetry and results in an angular-dependent 21-cm signal. The non-linear anisotropic density contrast, $\delta_{21}^{\rm nonlin}({\bf k},z)$,  is calculated from the real space isotropic signal   by  multiplying it by the factor $\left(dv_r/dr\right)^{-1}$. Because on the scales explored with our simulations the peculiar velocities are small compared to the Hubble flow, this factor can be approximated by $H^{-1}(z)(1+z)\left[1-\delta_{d_rv_r}({\bf x},z)\right]$  where $\delta_{d_rv_r}({\bf x},z)$ is the dimensionless velocity perturbation in real space calculated as the inverse Fourier transform of  $\delta_{d_rv_r}({\bf k},z) = -\mu_{{\bf k}}^2\delta({\bf k},z)$ \citep{Kaiser:1987, Bharadwaj:2004, Barkana:2005}.   This allows us to combine the linear peculiar velocity field with the other sources of 21-cm fluctuations in real space, non-linearly.  In other words,  the addition of the velocity field is non-linear even though its magnitude is calculated using linear theory (peculiar velocities are small compared to the Hubble flow, not compared to other perturbations). Finally, we take another Fourier transform in order to obtain the 21-cm fluctuation signal in k-space. We refer to this approach as our ``non-linear" case. As an illustration, in Figure \ref{fig:PS} we show power spectra for the astrophysical Case1 at $z = 19$ (close to the redshift of the absorption trough). The Figure shows the spherically averaged isotropic power spectrum as a function of $k$ (left) and the anisotropic two-dimensional power spectrum as a function of  $k_{\rm par} = \mu k$, projection of the wavenumber along the line of sight (more precisely, we plot the power spectrum as a function of the absolute value of $k_{\rm par}$), and $k_{\rm perp} = k\sqrt{1-\mu^2}$, perpendicular to the line of sight. 

It is useful to compare the non-linear case to the  fully linearized version of the anisotropic 21-cm signal  \citep[e.g.,][]{Barkana:2005}. In the linear regime, perturbations are additive and the anisotropic signal is given by
\begin{equation}
\delta_{21}^{\rm lin}({\bf k},z) = \delta_{21}^{\rm iso}({\bf k},z)- T_{21}\delta_{d_rv_r}({\bf k},z), 
\label{d21}
\end{equation}
where the minus sign arises because the velocity term is in denominator (Eq. \ref{Eq:tau}). We refer to this approach as the ``linear" case. In the linear regime $\delta_{21}^{\rm lin}({\bf k},z)$  can also be written as $T_{21}\left[W(k,z)+ \mu_{{\bf k}}^2\right]\delta({\bf k},z)$. We will make use of this property in Section \ref{Sec:Fitting}.

\subsection{Galaxies}
\label{Sec:gals}

Our second tracer is a mock galaxy survey calculated using the same simulations which were used to generate the 21-cm signals.   As described in Section \ref{sec:Mdata}, each halo of mass above the atomic cooling threshold hosts a galaxy with the stellar mass proportional to its halo mass. Therefore, the simulation readily provides a galaxy catalogue. The treatment is self-consistent in that the 21-cm signal and the mock galaxy survey are probing the same large-scale density field and the same realization of galaxies.

 In order to illustrate the technique,  we  first  consider a  simplified (``ideal")  survey where the galaxy density field is modelled as in Eq. \ref{dgal} assuming a fixed galaxy bias, $b = 5$. In this case, the effective number of galaxies in each simulated real-space cell is not necessarily an integer, as it is set by the cosmic mean abundance modified with respect to the local value of the large-scale density field.

Next, we generate a simple approximation of a more realistic galaxy field. We use the minimum mass of observed star forming halos, M$_{\rm min}$, as a proxy for telescope sensitivity. A low threshold, M$_{\rm min} = 10^7$ M$_\odot$, mimics observations with an extremely sensitive telescope that  can detect faint high-redshift galaxies formed in minihalos. However, in reality such sensitive surveys will most likely probe only a small part of the sky. To make the multi-tracer analysis between galaxy surveys and the SKA (or similar) possible, large survey areas are required implying that the sensitivity will most likely be compromised.  Therefore, we also  consider higher values of   M$_{\rm min} = 10^8$ M$_\odot$ and M$_{\rm min} = 10^9$ M$_\odot$. The minimum observable halo mass can be related to stellar mass (as described in Section \ref{sec:Mdata}). 
In the redshift range $z = 6-15$ and assuming  cosmic mean value of the relative velocity between dark matter and gas,  we find that   M$_{\rm min} = 10^8$ M$_\odot$ corresponds to M$_* \approx  7 \times 10^5$  M$_\odot$ and M$_{\rm min} = 10^9$ M$_\odot$ corresponds to M$_* \approx 7 \times 10^6$  M$_\odot$.  Relating the stellar masses directly to telescope sensitivity requires additional model for dust attenuation and is out of the scope of this paper.


To ensure that our galaxy distribution is realistic, we also make sure that  the number of galaxies in any given cell,  $n_{\rm cell}$, is integer. We calculate $n_{\rm cell}$ by drawing it from a Poisson distribution with the  mean value given by the mean expected number of galaxies\footnote{ Importantly, the number of galaxies in each 3 Mpc cell is derived using the same statistics of collapsed objects that is used to calculate the 21-cm signal. This model is self-consistent in that the 21-cm signal and the mock galaxy survey are indeed probing the same realization of galaxies.}
 $\bar n_{\rm cell} = V_{\rm cell} <n>\left[1+b(z) \delta_{\rm cell}\right]$. Here $V_{\rm cell} = 27$ Mpc$^3$ is the comoving volume of each cell in our simulation, $<n>$ is the mean cosmic  abundance of halos above ${\rm M}_{\rm min}$,  $\delta_{\rm cell}$ is the mean overdensity of the cell, and $b(z)$ is the redshift-dependent galaxy bias which we calculate from the collapsed fraction,  $b(z)  = d\log f_{\rm coll}({\rm M}_{\rm h}>{\rm M}_{\rm min})/d\delta$.  The noisy biased isotropic galaxy density field is, thus,  $\delta_{\rm gal}({\bf x},z) = n_{\rm cell}\left[V_{\rm cell}<n>\right]^{-1}-1$.  Finally, anisotropy is added by applying the Kaiser effect  as explained in Section \ref{Sec:anis} (in the linear case).

\begin{figure*}
\includegraphics[width=3.4in]{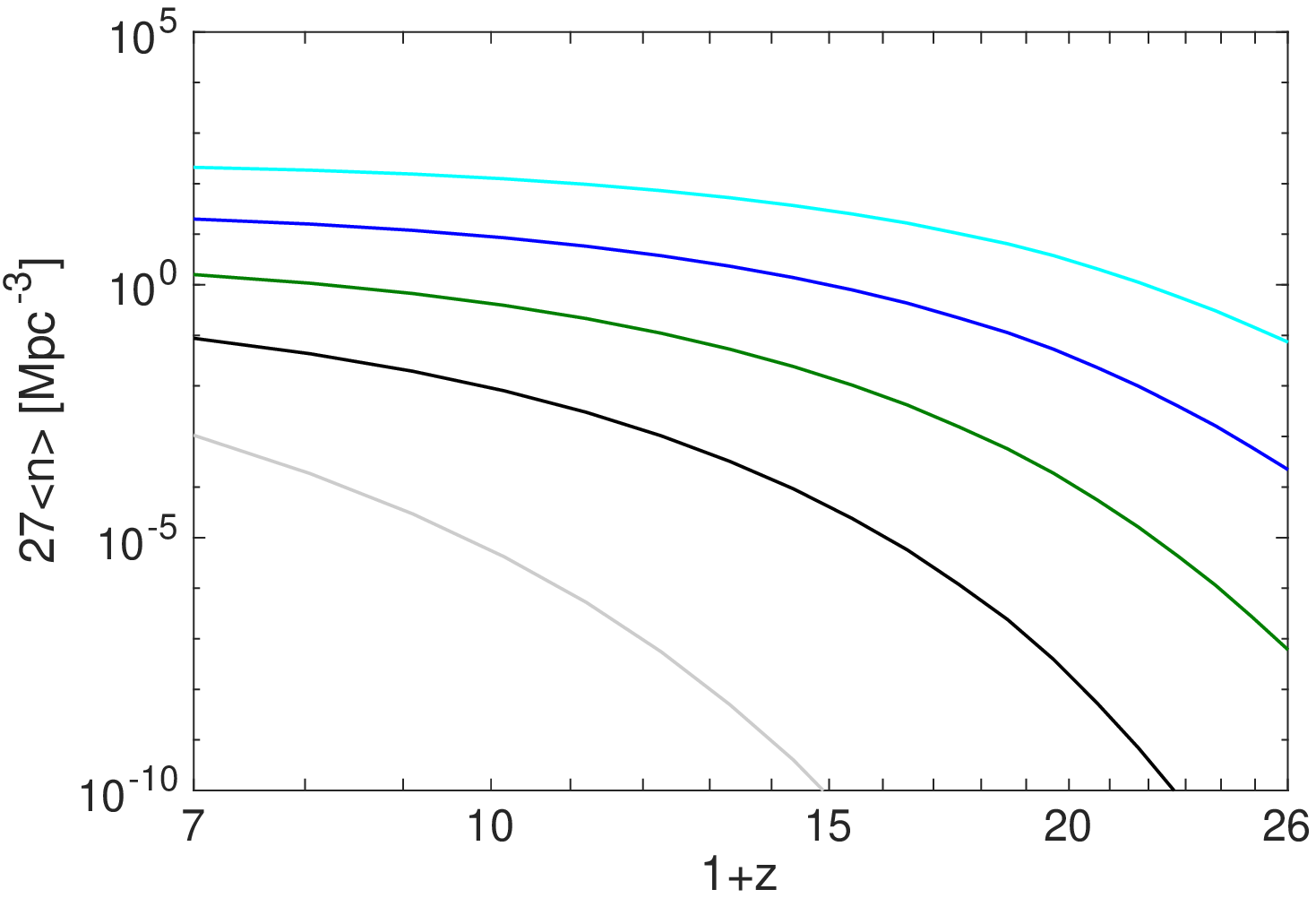}\includegraphics[width=3.4in]{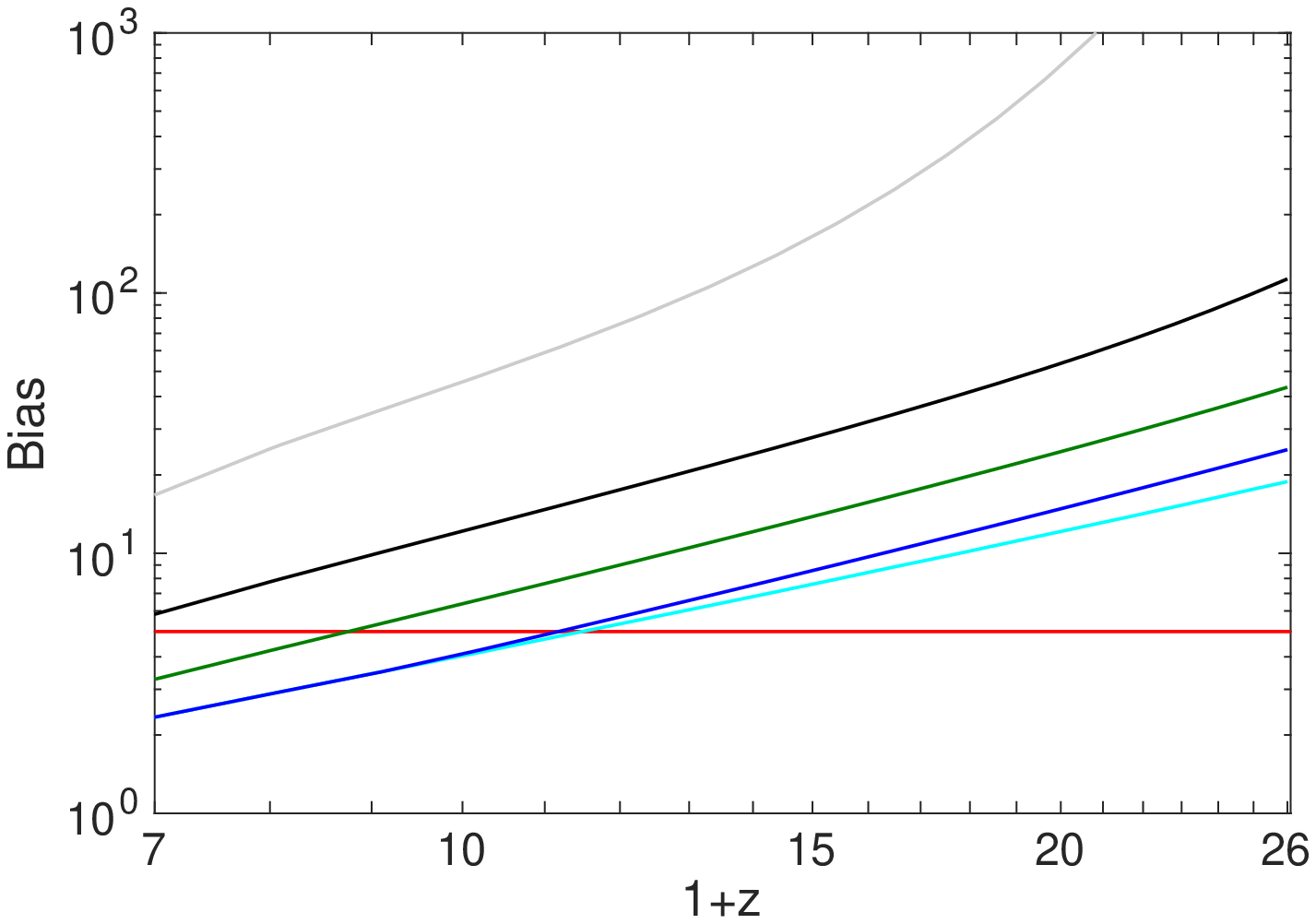}
\caption{Left: Cosmic mean number of halos in a cell, $\bar n_{\rm cell}$. We consider  M$_{\rm min} = 10^7$ M$_{\odot}$ (cyan), M$_{\rm min} = 10^8$ M$_{\odot}$ (blue), M$_{\rm min} = 10^9$ M$_{\odot}$ (green), M$_{\rm min} = 10^{10}$ M$_{\odot}$ (black), M$_{\rm min} = 10^{11}$ M$_{\odot}$ (grey). Right: Redshift-dependent galaxy bias $b(z)$.  The color code is the same as in the left panel. The scale-independent bias $b = 5$ (red horizontal line) is added for comparison. }
\label{fig:bn}
\end{figure*}

Examples of the cosmic mean number of halos in a cell and the redshift-dependent galaxy bias $b(z)$ are shown in Figure \ref{fig:bn}   as a function of  redshift for several choices of ${\rm M}_{\rm min}$. The higher is the cutoff scale  the lower is the number of star-forming halos and the higher is the bias \citep[although our calculated bias is somewhat higher than the measurements at $z\sim 6$, see][]{Harikane:2016, Hatfield:2018}. In contrast, the anisotropy term does not depend on the cutoff scale. Therefore, for higher  M$_{\rm min}$ the anisotropy is effectively weaker. This is the reason why (as we will see in Section \ref{Sec:Pois}) the method, which we explain in detail in Section \ref{Sec:Fitting}, does not work for noisy galaxy samples with high  M$_{\rm min}$.

\section{Parameter Estimation}
\label{Sec:Fitting}

In this Section we show how the ratio of the two density fields,  $\delta_{21}({\bf k},z)$ and $ \delta_{\rm gal}({\bf k},z)$, assuming an idealized galaxy survey, can be used to measure the global (sky-averaged)  21-cm signal, the window function and properties of the galaxy population (bias). Consider, first, the ratio in the linear regime
\begin{equation}
R^{\rm lin}_{{\bf k}}(z) \equiv \frac{\delta_{21}^{\rm lin}({\bf k},z)}{\delta_{\rm gal}({\bf k},z)}.
\end{equation}
 Because, as can be seen from Eqs.  \ref{dgal} and \ref{d21}, in the linear approximation both of these fields are proportional to $ \delta({\bf k},z)$, its contribution cancels out and  the ratio can be written as a deterministic function\footnote{More precisely, at every redshift Eq. \ref{Eq:Rk} represents a set of $N_k$ functions, where $N_k$ is the number of samples of $k$.}  of  $X \equiv \mu^2_{\bf k}$ parameterized by   $T_{21}(z)$, $W(k,z)$ and $b$
\begin{equation}
R^{\rm lin}_{{\bf k}}(z)  = \frac{T_{21}(z)\left[W(k,z)+\mu^2_{\bf k}\right]}{b+\mu^2_{\bf k}}.
\label{Eq:Rk}
\end{equation}
  Here we are only interested in high-redshift applications of the method ($6<z<26$, matter dominated Universe) and set $f=1$.

Our goal is to extract $T_{21}(z)$, $W(k,z)$ and $b$ from the ratio of  the simulated fields $\delta_{21}^{\rm lin}({\bf k},z)$ and $\delta_{\rm gal}({\bf k},z)$.  To this end at every $k$ and $z$ we bin the ratio into $N_X$ bins along  $X$, and fit it  by a function of the form 
\begin{equation}
F_k(X) = \frac{A_k+BX}{C+X}.
\label{Eq:Fit}
\end{equation}
We measure the values of $A_k$, $B$ and $C$, which are then used to estimate the astrophysical quantities:
\begin{align}
\hat b &= C,\\
\hat T_{21}(z) &= B,\\
\hat W(k,z) &= A_k/B.
\label{Eq:fit2}
\end{align}
At the end of this procedure (details are discussed in   Appendix \ref{App1}) we have the  best fit values of $\hat b(z)$, $\hat T_{21}(z)$ and $\hat W(k,z)$ together with $1-\sigma$ statistical errors in each parameter  (calculated using standard error propagation analysis). Note that without the anisotropic terms, only the
degenerate combination $T_{21}(z)W(k,z)/b$ can be measured.

\subsection{Accuracy}
\begin{figure*}
\includegraphics[width=3.4in]{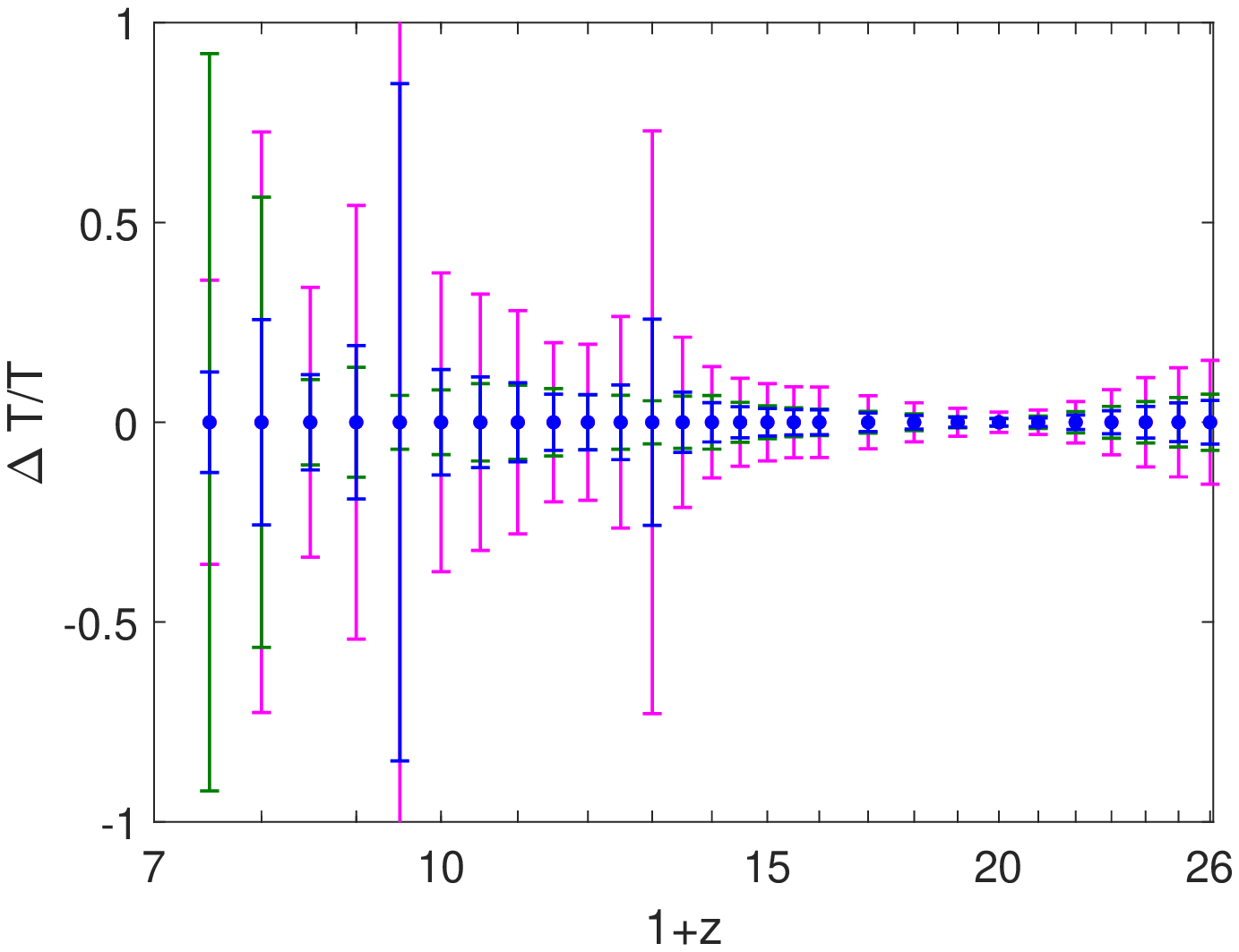}\includegraphics[width=3.4in]{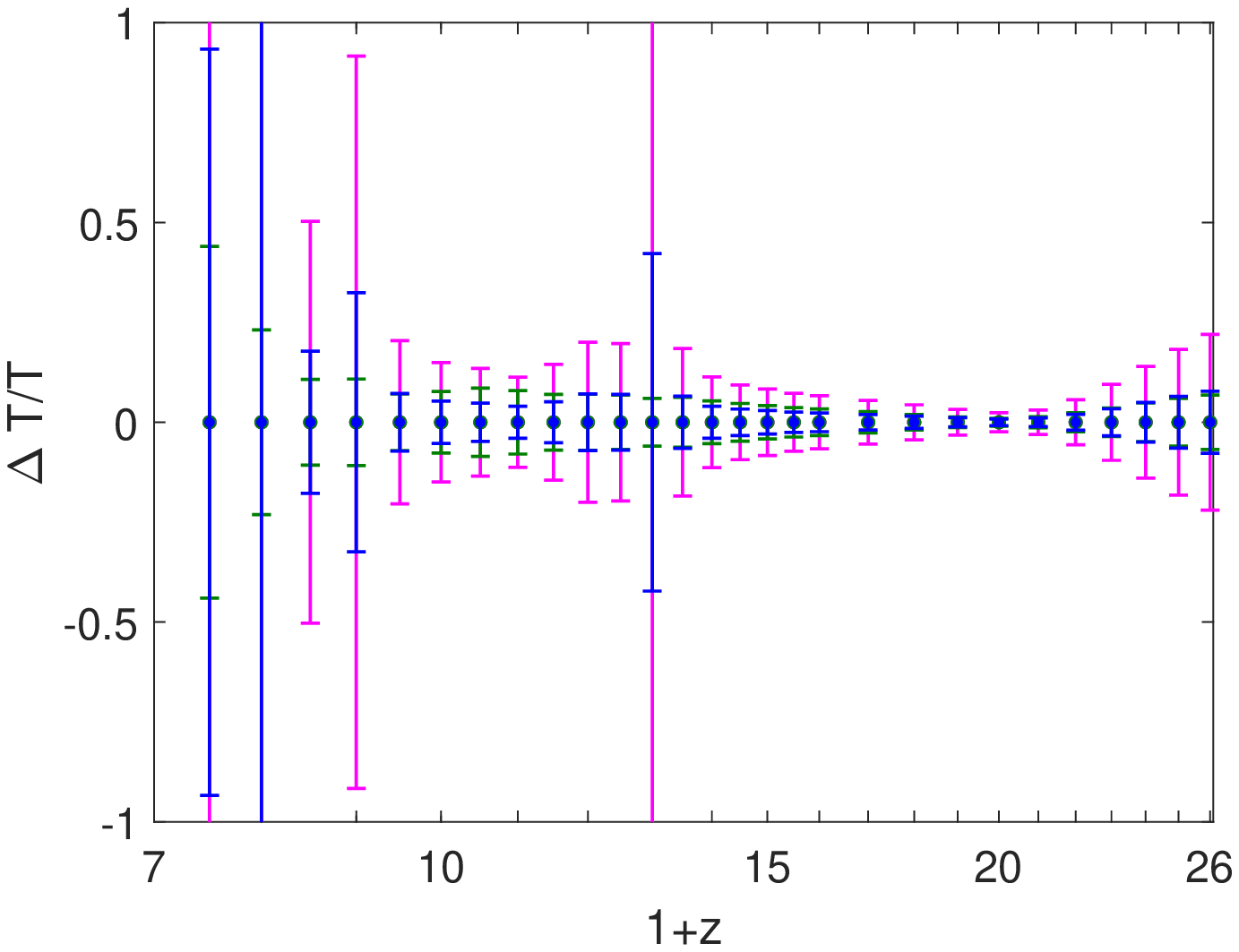}
\caption{Statistical  dimensionless errors in the brightness temperature  ($\Delta T/T$) for  boxes with $128^3$ cells ($\Delta_{128}$, magenta) and  $256^3$ cells ($\Delta_{256}$, green). The scaled version  $ \tilde \Delta_{256} = \Delta_{128}\left(128/256\right)^{3/2}$ is also shown (blue). Left and right panels show the results drawn from different realizations of initial conditions. The results are shown for the astrophysical Case2   for which the heating transition happens at $z = 12.1$ and assuming an idealized galaxy survey. We note that while this figure only shows statistical errors, Fig. \ref{Fig:T2} below shows the systematic errors as well.}
\label{Fig:CB}
\end{figure*}

Our main results (discussed in Section \ref{Sec:Res}) are derived from simulations of comoving volume  768$^3$ Mpc$^3$ (consisting of 256$^3$ cells). This volume is  close to the projected field of view of the SKA, $\sim 5$ degrees at 100 MHz, which at $z=9$ corresponds to $\sim 823$ comoving Mpc. However, a few survey modes with the SKA are being  discussed which would cover a larger area by surveying multiple fields of view, either 5 (deep), 50 (medium) or 500 (shallow) fields. 
Because simulating larger boxes than 768$^3$ Mpc$^3$  is not feasible at the moment, we explore the scaling of the results by down-sizing the simulations. We compare with the results from smaller simulations of  384$^3$ Mpc$^3$ performed at the same resolution as the larger volumes. It is expected that, if the errors are dominated by statistical errors, decreasing the size of the simulated volume should degrade the parameter estimation. In this case, decreasing the  number of statistically independent modes in the data cube from $N^3$ to $M^3$ is expected to increase  the error by  a factor of  $\left(N/M\right)^{3/2}$. Our results closely follow this trend. The errors in the brightness temperature (i.e., the global 21-cm signal) as a function of redshift are demonstrated in Figure \ref{Fig:CB} for two independent sets of initial conditions (left and right panels)  for a box with $128^3$ modes ($\Delta_{128}$, magenta) and  $256^3$ modes ($\Delta_{256}$, green); in addition, we show the scaled version of $\Delta_{128}$ multiplied by the numerical factor of $\left(128/256\right)^{3/2}$ ($\tilde \Delta_{256}$, blue). The value of  $\Delta_{256}$ is very close to  $\tilde \Delta_{256}$ at the majority of the considered redshifts indicating that the errors are of statistical origin at most redshifts. Therefore, the larger the survey, the more reliable reconstruction can be achieved. 
 Comparing the right and the left panels we see that the size of the errorbars only mildly depends on the specific realization of the initial density field. Provided that the fluctuations are detected with high enough significance, we conclude that the SKA shallow survey should be the most effective survey for extracting  the global 21-cm signal (assuming  perfect extraction of the 21-cm field and  that there exists a deep enough large-scale survey or another tracer to accompany the SKA measurement). 

 Statistical errors in galaxy bias and $W(k,z)$ (e.g., error bars in the right panel of Figure \ref{Fig:T2} and in Figure \ref{fig:WL1}) are also calculated via standard error propagation analysis. Because at each redshift the fitting is done simultaneously for all the $N_k+2$ parameters -- $N_k$ samples of $W(k,z)$, and one sample of each $b$ and $T_{21}(z)$ -- the strength of the  21-cm signal affects measurement of the bias and the window function. 
The quality of the fit is poor (and statistical errors are high) for low amplitude signals, which includes the redshift of the heating transition at which $\hat T_{21}(z)$ is close to zero, and low redshifts where the signal vanishes due to reionization. On the other hand, the fitting procedure performs well when the signal is strong, with the errors dropping to minimum at around $z\sim20$ where the global signal reaches its maximum. Furthermore, for $W(k,z)$ the errors are larger at lower $k$-modes (corresponding to larger spacial scales) because then the number of statistically independent modes is smaller. 
Quantitative discussion on the errors can be found in the next Section.

 In this paper we assume perfect foreground removal/avoidance at wavenumbers between  $k\sim 2\pi/$[box size] (corresponding to the largest accessible scale) and 1 Mpc$^{-1}$. However, our method also works if the lowest modes are lost in the process of foreground subtraction. We test the robustness of the method by comparing the results obtained using wavenumbers in the  $0.01-1$ Mpc$^{-1}$ range  and  $0.05-1$  Mpc$^{-1}$ range finding a very minor effect (see Figure \ref{Fig:app} in the Appendix).

 \section{Results}
 \label{Sec:Res}
Even though Eq. \ref{Eq:Fit} was derived in the linear regime, here we use it to fit the ratio of the non-linear 21-cm field and an ideal (Section \ref{sec:id}) as well as a realistic (Section \ref{Sec:Pois}) galaxy sample.  As we will see below, the fitting procedure works well when Eq. \ref{Eq:Rk} is  a good representation of the data, i.e., when both the non-linearity and galaxy bias are small. The method is expected to perform poorly  for highly biased galaxy sample or when non-linearity is strong. In these cases systematic errors are expected. 
 
 \subsection{Ideal Galaxy Surveys}
 \label{sec:id} 
 
\begin{figure*}
\includegraphics[width=3.4in]{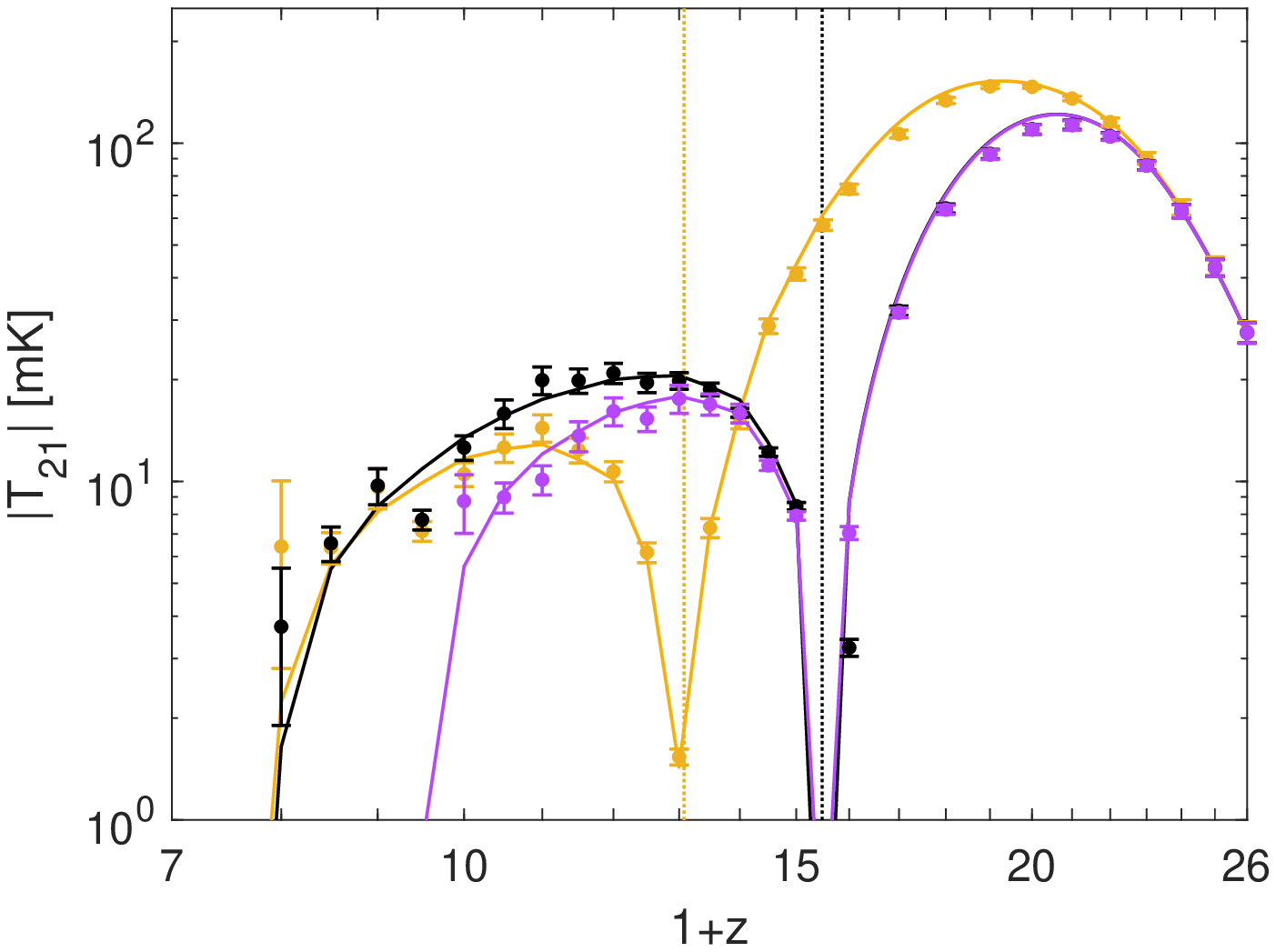}\includegraphics[width=3.4in]{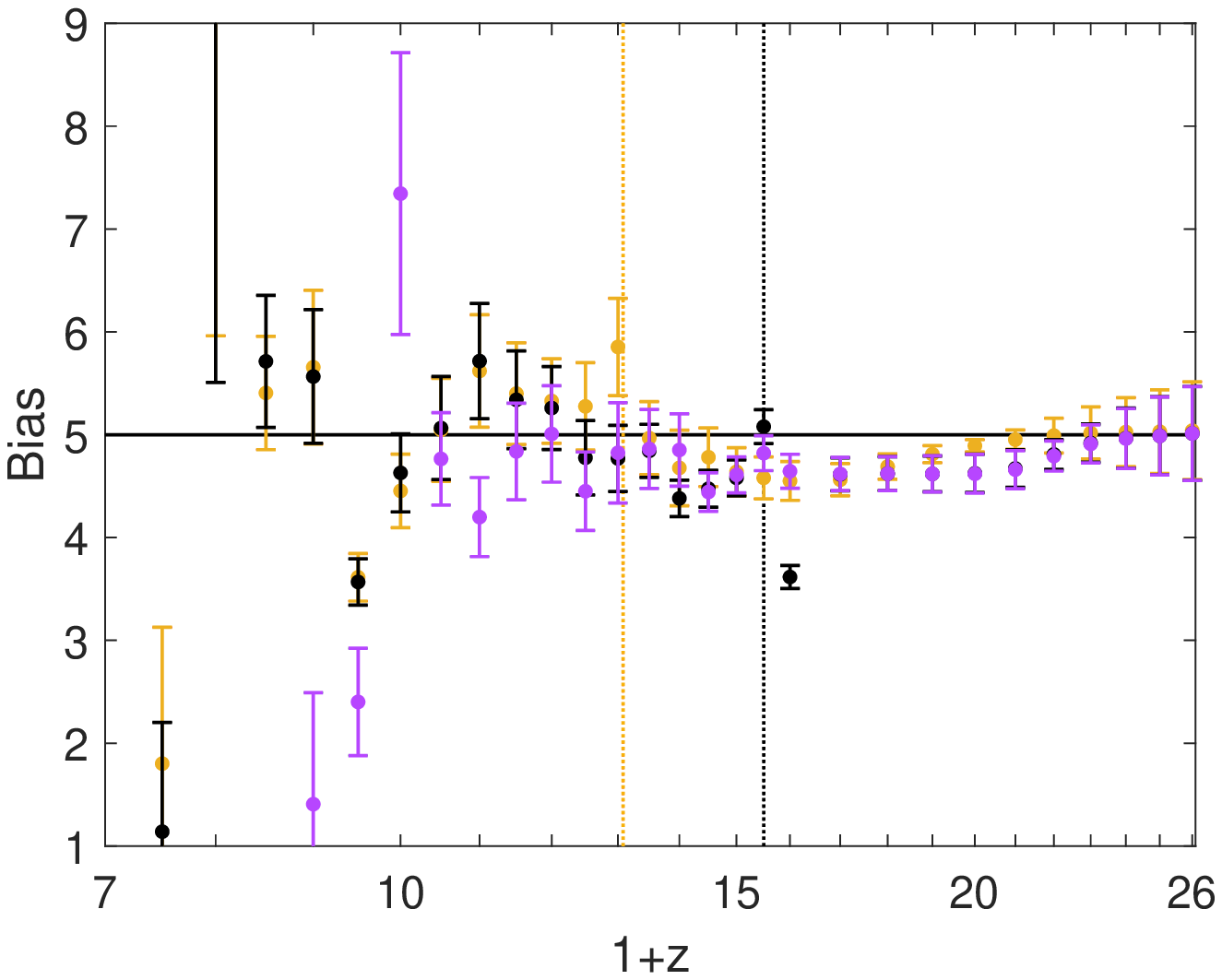}
\caption{Left: The amplitude of the simulated global 21-cm signals (solid lines, log scale) and measured $|\hat T_{21}|$ (data points)  extracted using the multi-tracer technique from the data with velocities added in real space (the non-linear case). We show   Case1 (black),  Case2  (orange),   Case3 (purple).   Each measurement is a result of fitting, extracted from one 768$^3$ Mpc$^3$ box, where no Poisson noise was added. The errorbars indicate the 1-sigma error of nonlinear regression calculated from the joint
fitting of the temperature, window function and bias. Vertical lines indicate the redshift of heating transition in each case.
Right: Expected bias $b=5$ (horizontal line) and measured bias $\hat b$ (points) extracted using the multi-tracer technique. Same color code as on the left.}
\label{Fig:T2}
\end{figure*}

Assuming an ideal galaxy survey, we extract the mean values and $1-\sigma$ error bars of the brightness temperature and galaxy bias and show them in Figure \ref{Fig:T2}. Original values of these parameters (the global signal directly measured from the simulated field $T_{21}({\bf x}, z)$ and the constant galaxy bias $b=5$) are shown for comparison.  The results are shown over a wide range of redshifts and for different sets of astrophysical parameters (Case1,  Case2  and   Case3 as discussed above). 

As is evident from Figures  \ref{Fig:CB} and \ref{Fig:T2},  errors in the global signal are of order of ten per cent at high redshifts where the signal is very strong and non-linearity is weak ($9$\% at $z = 17$ for Case1, including both the statistical error and the systematic offset). One exception is $z_h$. At this relatively high redshift the global 21-cm signal is close to zero and the errors diverge. At lower redshifts the reconstruction errors  are higher (increasing to few tens percent, $21\%$ at $z=8$ for Case1) because  the non-linearity is strong and the signal is weak due to the ongoing reionization and saturated heating.  Specifically, close to $z_{\rm eor}$  the errors are of order $\sim 100\%$ as the 21-cm signal approaches zero, preventing efficient parameter estimation.  As is evident from the left panel of Figure  \ref{Fig:T2}, throughout the cosmic history the errors  in the reconstructed brightness temperature are much smaller than the variation of the signal introduced by the change in the astrophysical parameters (namely, X-ray SED and $\tau$, both  varied within margins allowed by existing observations). This indicates that the multi-tracer method could potentially yield constraints on the astrophysical parameters from the observed data, especially if wide-field surveys with the SKA are used. In a similar manner, the error in the galaxy bias  (shown in the right panel of Figure \ref{Fig:T2}) is of order ten per cent at high redshifts where the method works well (7\% at $z = 17$ for Case1, including both the statistical error and the systematic offset), increasing  to few tens of per cent at lower redshifts ($\sim 17$\%  at $z= 8$ for Case1).

\begin{figure*}
\includegraphics[width=3.4in]{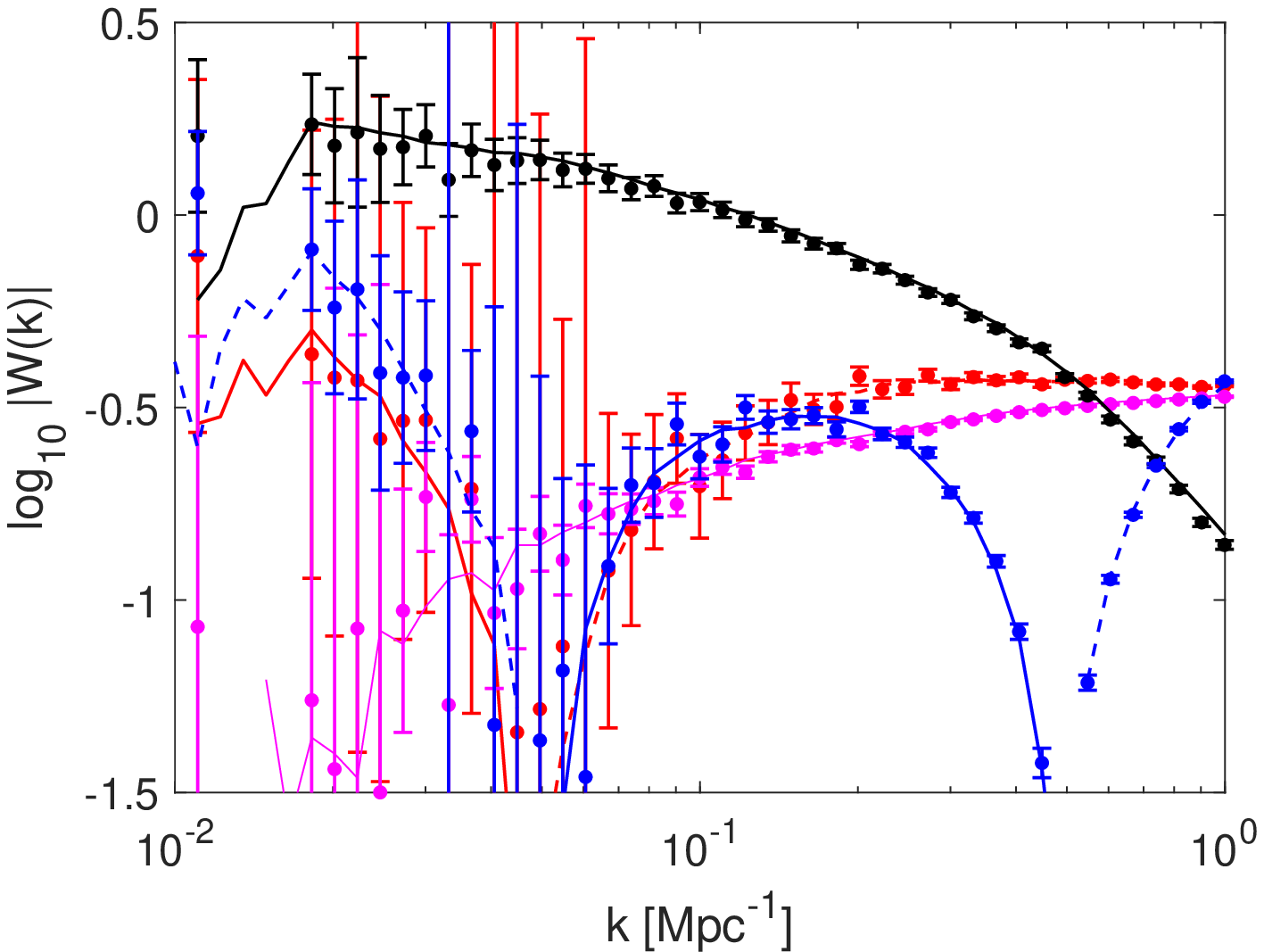}\includegraphics[width=3.4in]{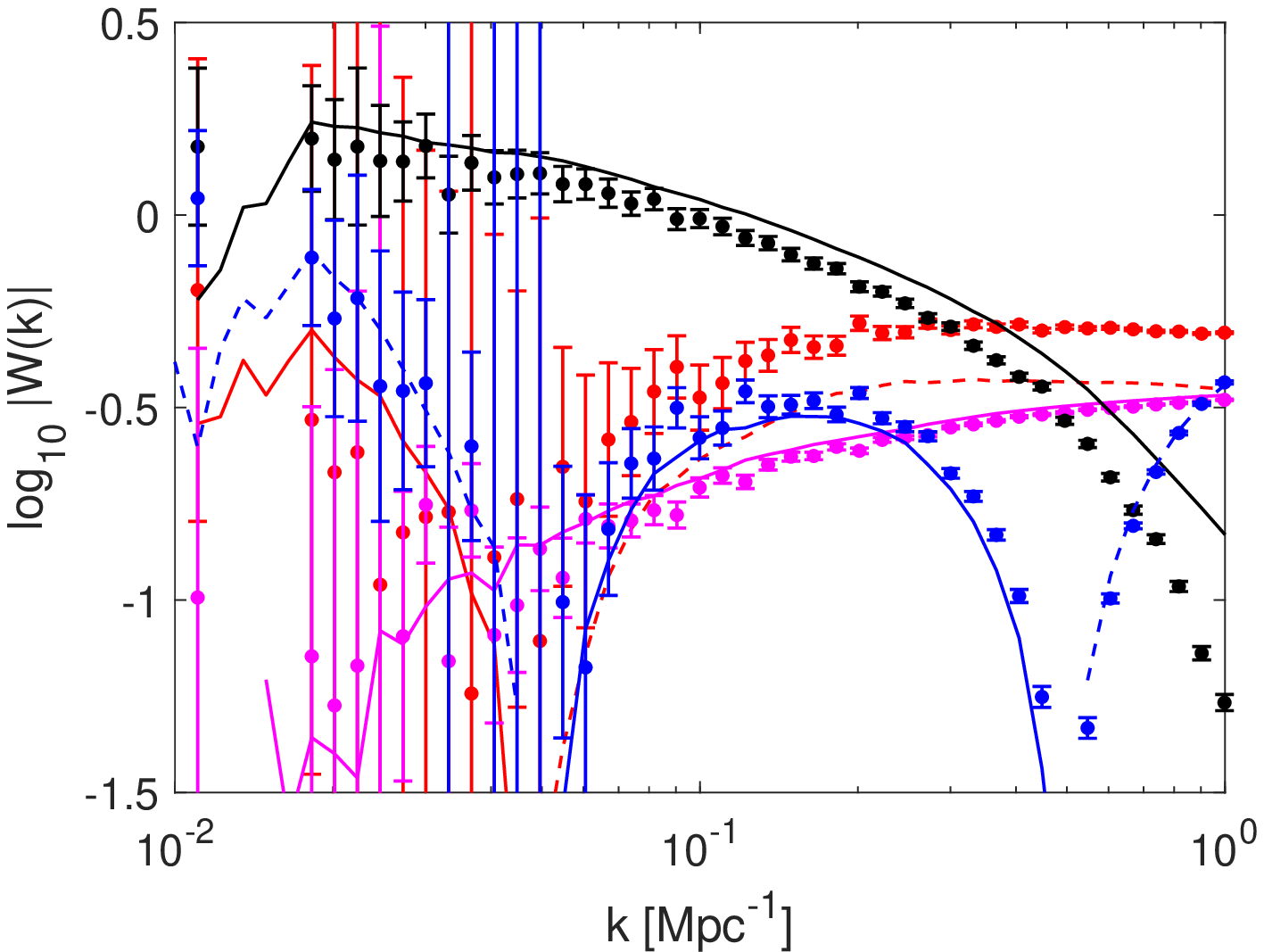}
\caption{Logarithm (base 10) of the absolute value of the window function is shown as a function of wave number with in the linear  (left) and non-linear (right) cases.    The theoretical  $W(k,z)$ is shown with solid (positive value) and dashed (negative value). Astrophysical models are shown for  Case1  and   Case2  at the redshift of the global maximum/minimum of $T_{21}$:   $z = 12/20$  for  Case1 (black/blue) and  $z =10/18$ for Case2 (red/magenta). The reconstructed $\hat W(k,z)$ is marked with points of the corresponding colors. $\hat W(k,z)$  has the correct sign everywhere (matching $W(k,z)$). We also show $1-\sigma$ fitting error bars for each measurement. }
\label{fig:WL1} 
\end{figure*}

The window function of the dominant radiative source can also be measured. The estimates of $\hat W(k,z)$ as a function of $k$ are shown in Figure \ref{fig:WL1} at two characteristic redshifts corresponding to the absorption trough of the 21-cm signal and the emission peak.  The results are shown for the simulated Cases 1 \& 2.  We compare the expectation value $\hat W(k,z)$ with the theoretical value $W(k,z)$  computed using  Eq. \ref{Eq:iso} where $\delta_{\rm iso}$ is the isotropic density which we  measured directly from our simulations (prior to adding the velocity effects). The results are shown for the linear (left panel) and non-linear (right panel)  velocity perturbations.

In the case of linear theory   the reconstructed window function is in an excellent agreement with the theoretical curves for all the considered astrophysical scenarios and at all the explored redshifts. The errors are at sub-percent level at large $k$ where the number of independent samples per bin is large. At lower $k$ the errors are much larger because of the small number of statistically independent samples (showing that in this case we do not  evade the cosmic variance problem). 

In the non-linear case, Eq. \ref{Eq:Rk} is not expected to be in a perfect agreement with the simulated data and degradation in the quality of the fit (Eq. \ref{Eq:Fit}) is expected. Nevertheless, there is a good agreement between the theory and the measurement at high reshifts where non-linearity is small, the effect of X-rays is the strongest  and the signal itself ($|T_{21}|$) is the strongest. Specifically, at the redshift of deepest absorption  (blue/magenta  curve for Case1/2) both cases show a good agreement and small fitting errors at high $k$.  The reconstruction is slightly better for the hard SED (Case2) than for the soft SED (Case1) owing to the $\sim 30$\% deeper  absorption trough. At lower redshifts, although the shape of the window function can still be recognized and the sign of $\hat W(k,z)$ is always correct (same as of $W(k,z)$), there is an offset between the theoretical and the measured window functions  because the effect of  X-rays is  nearly saturated, non-linearity is large and the signal itself is low.  For example, at the redshift of the emission peak  (red and black lines in Figure \ref{fig:WL1}) the offset is above 50\%.

\begin{figure*}
\includegraphics[width=3.4in]{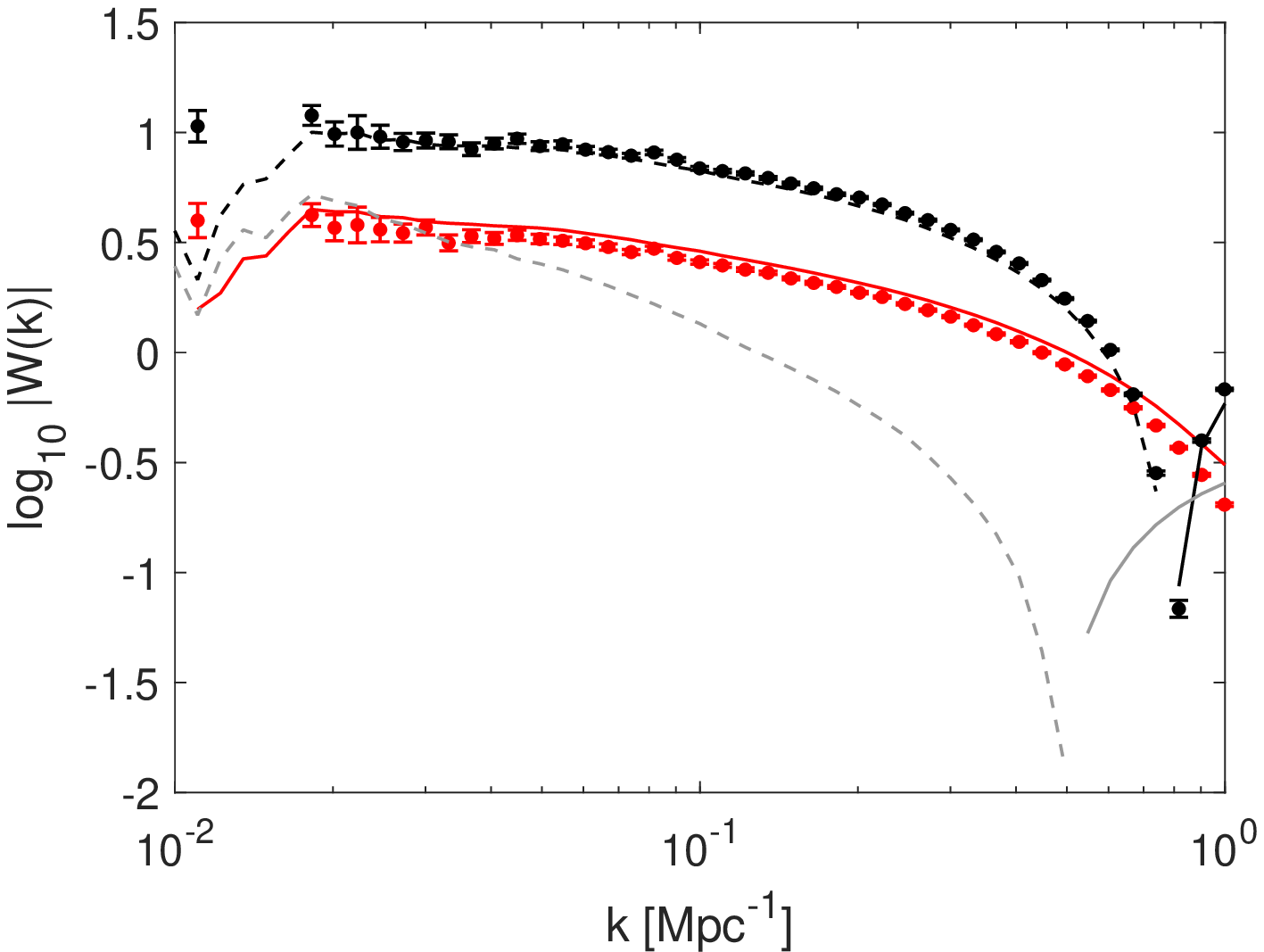}\includegraphics[width=3.4in]{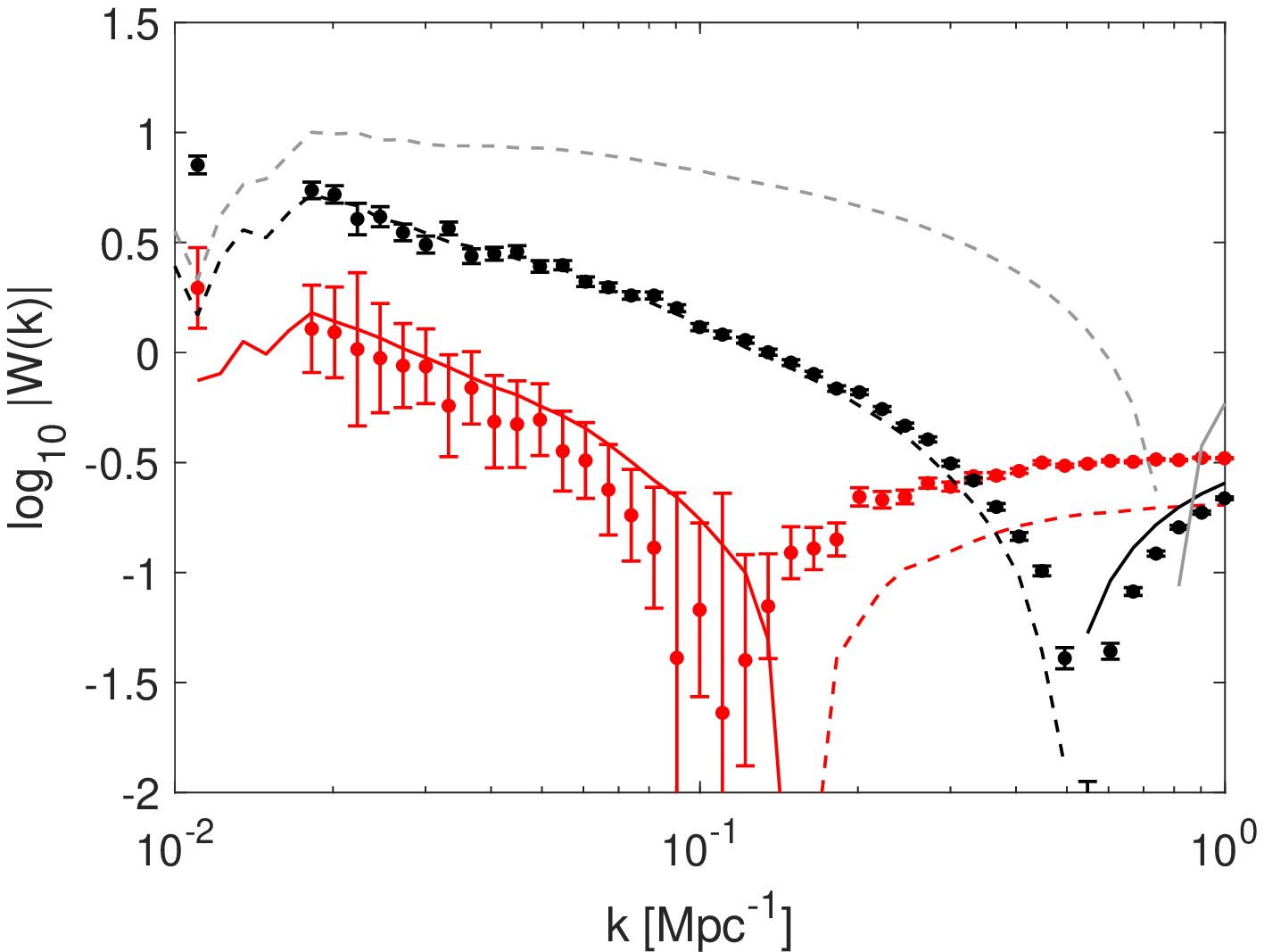}
\caption{We  compare the  window function  $\hat W(k,z)$ (logarithm, base 10, of the absolute value is shown) measured in the non-linear case (data points) to the theoretical window function $W(k,z)$ (curves) around the heating transition for  Case1   (soft X-ray SED, left) and  Case2  (hard X-ray SED, right). Solid lines show the positive value of $W(k,z)$, dashed - negative value. The fitted $\hat W(k,z)$ points have the correct signs (matching the $W(k,z)$ curves) in every case. The data is shown at  $z_h-1.5$ (red), $z_h+1.5$ (black). The grey curves on each panel show the highest redshift curve from the opposite panel. }
\label{Fig:HT}
\end{figure*}

Around the heating transition (when temperature fluctuations dominate the 21-cm power
spectrum), the slope of $W(k,z)$ is directly related  to the hardness of the X-ray SED \citep{Fialkov:2015}. If the spectrum is soft, heating happens predominantly on small spatial scales  and the window function in the Fourier space is steep.  On the other hand, energy is distributed over larger scales in the case of a hard SED and the window function is much flatter in the harmonic space.   In Figure \ref{Fig:HT} we demonstrate the variation of the window function with  X-ray SED and find a good agreement with our earlier results \citep{Fialkov:2015}. Therefore, if observed, the shape of $W(k,z)$ can be used to determine the SED of X-ray sources.

\subsection{Realistic Galaxy Surveys}
\label{Sec:Pois}

\begin{figure*}
\includegraphics[width=3.4in]{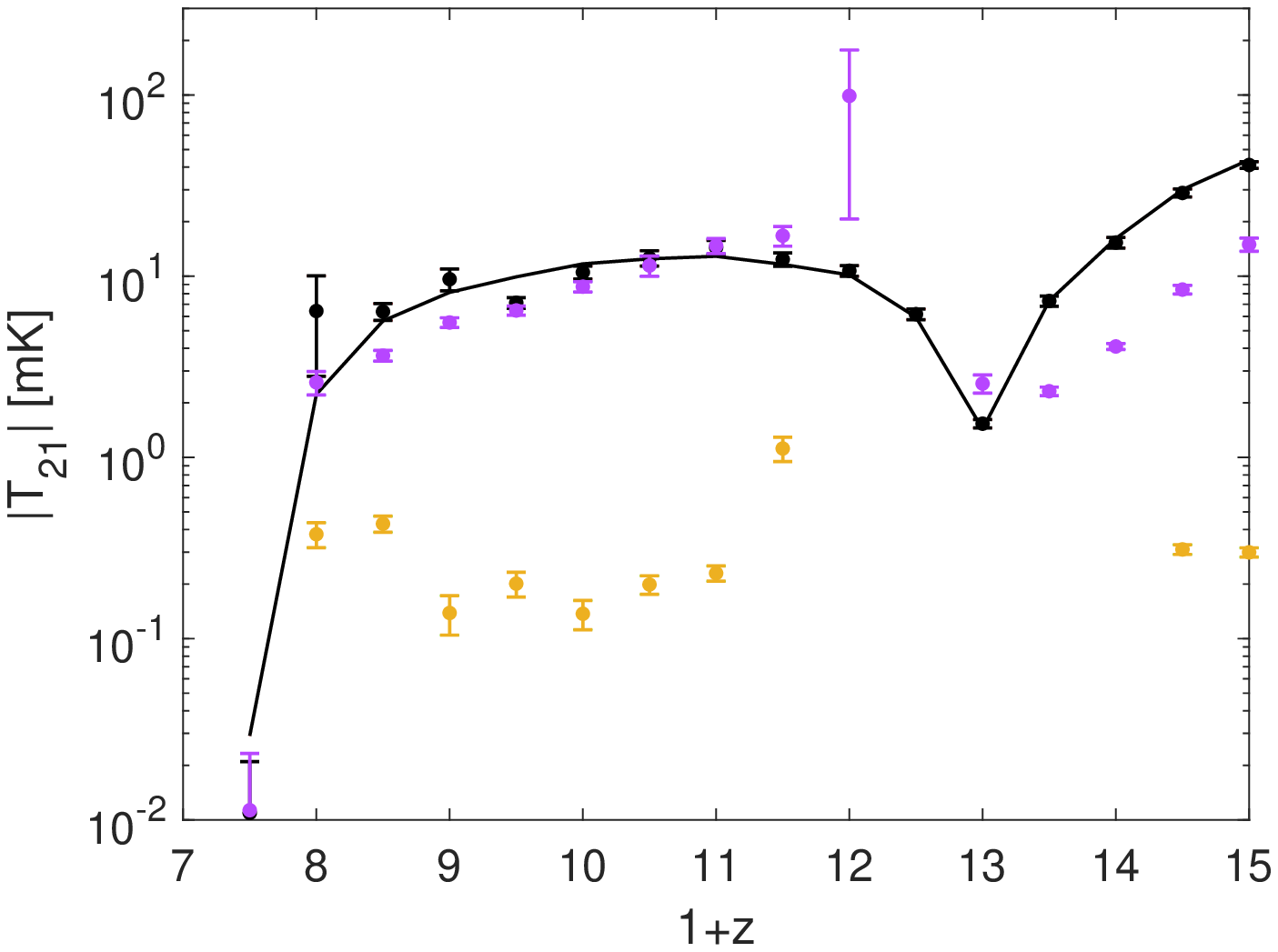}\includegraphics[width=3.4in]{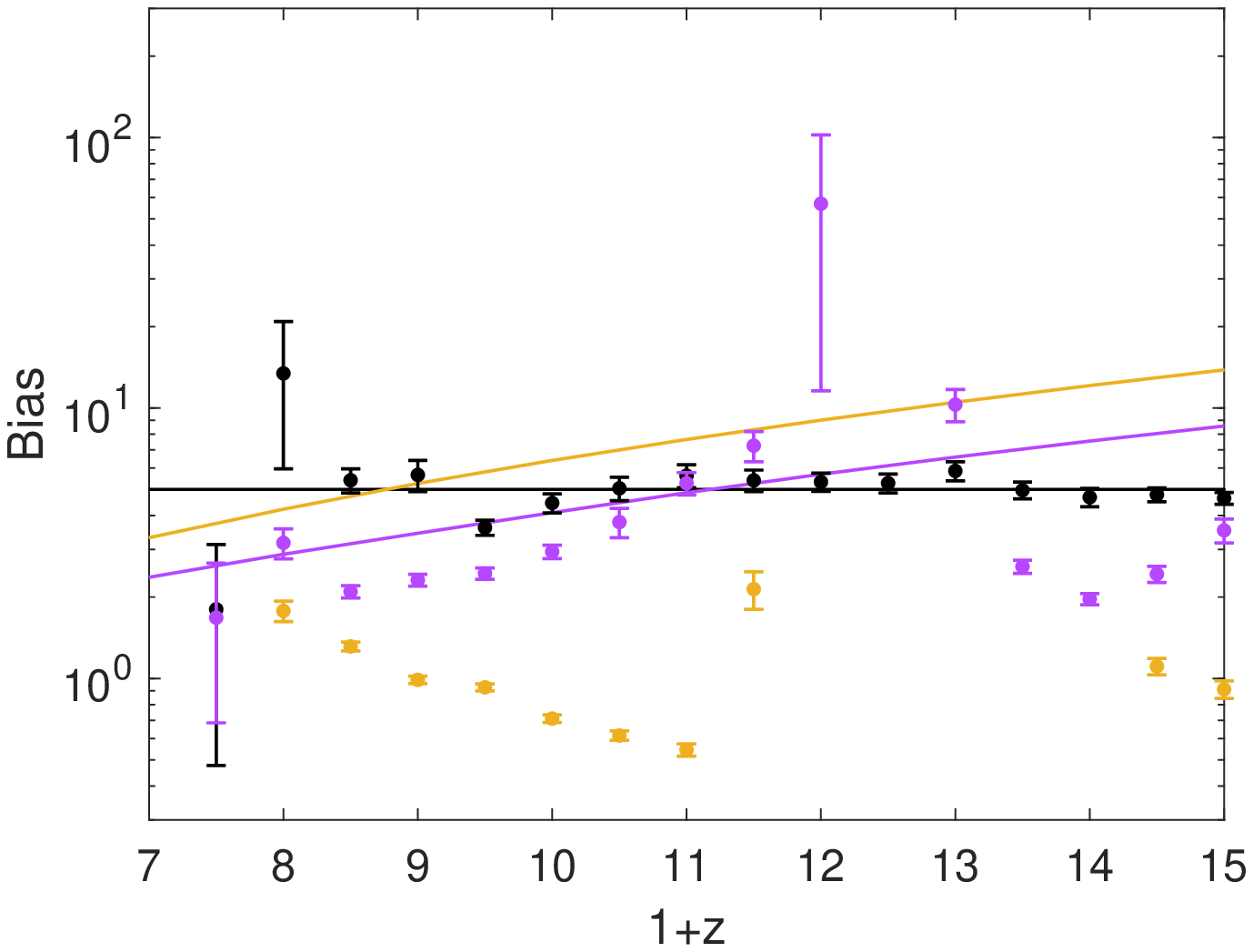}
\includegraphics[width=3.4in]{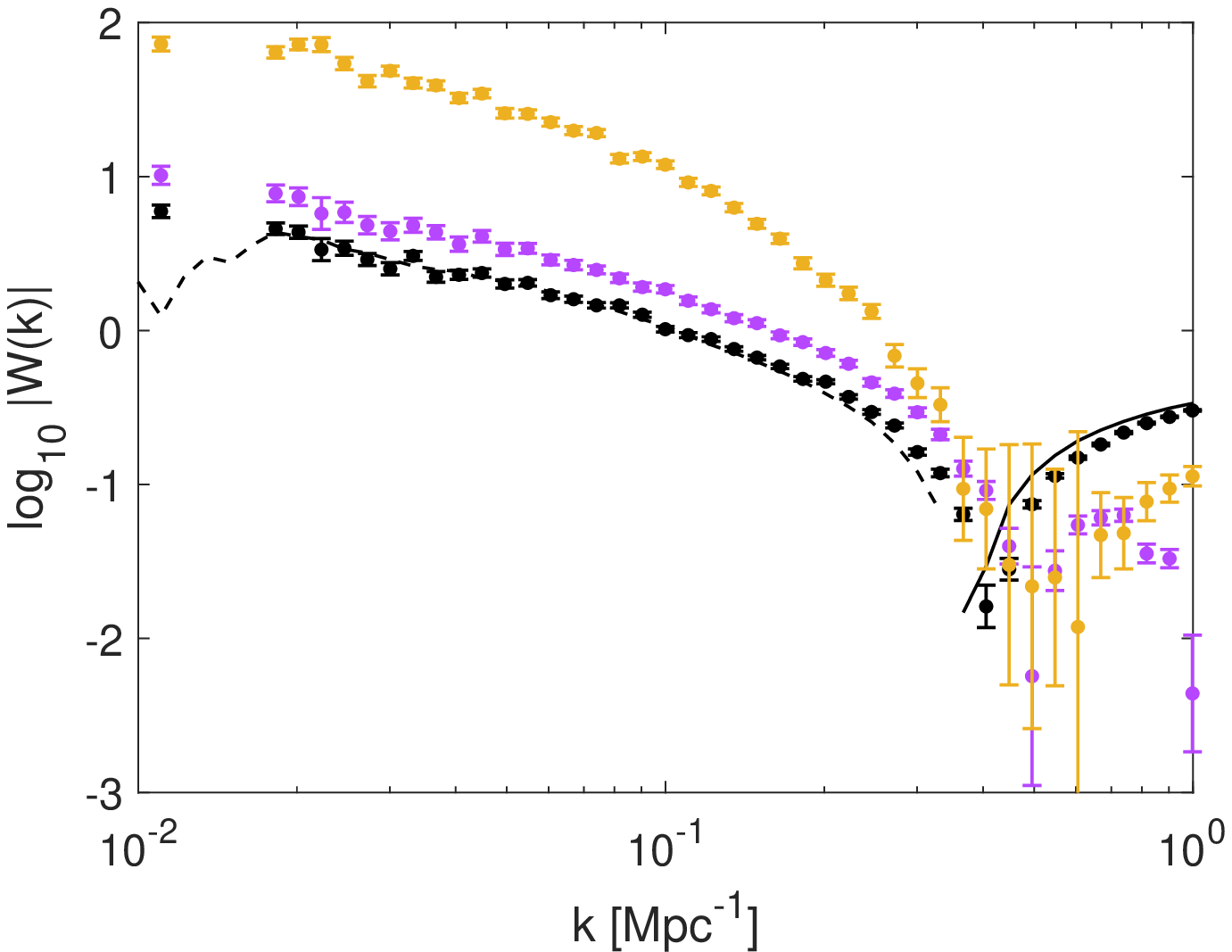}\includegraphics[width=3.4in]{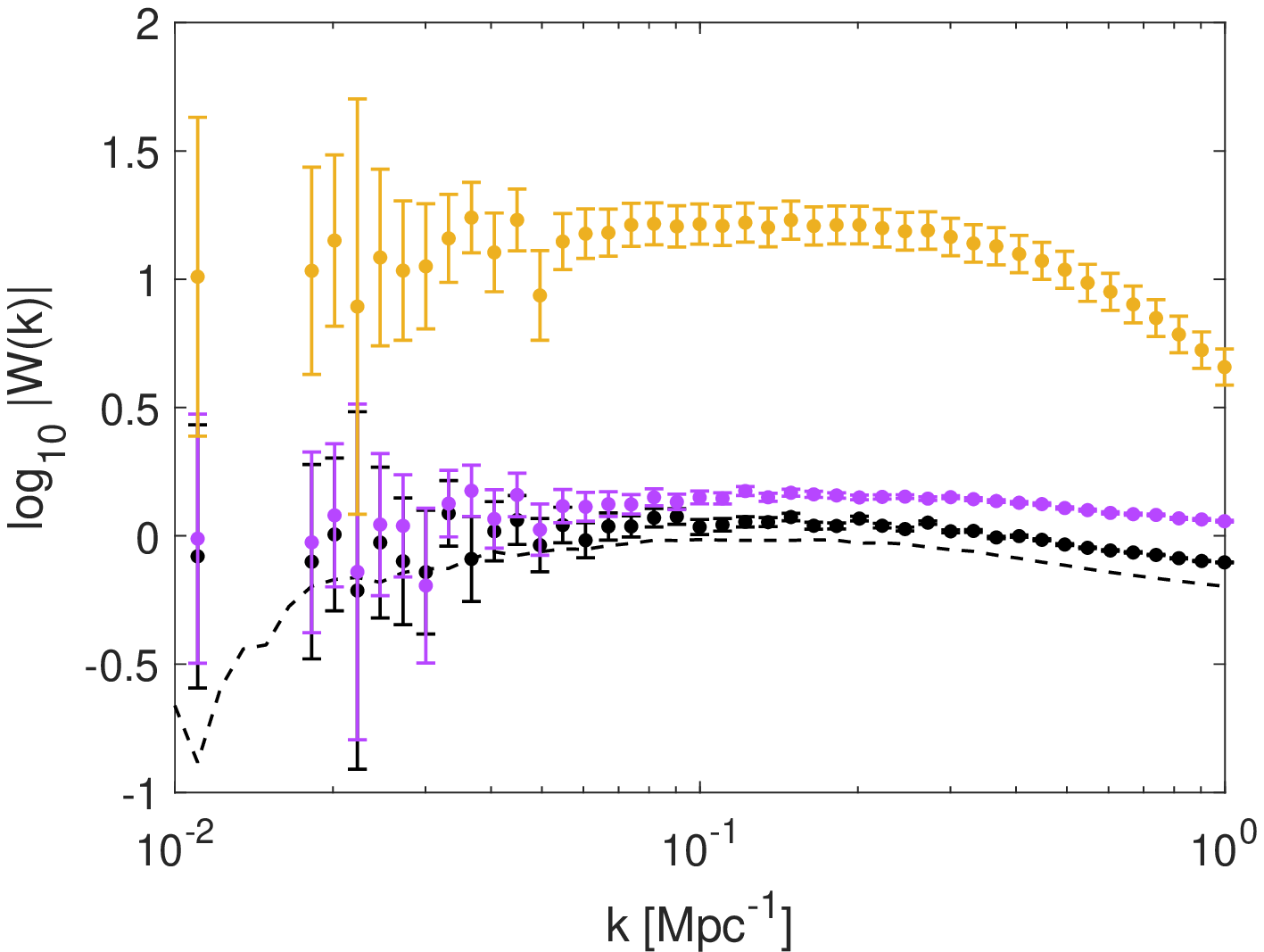}
\caption{Top Left: Global signal without Poisson noise (black) and with Poisson noise for the cutoff masses of M$_{\rm min}=10^8$ M$_\odot$ (purple) and M$_{\rm min} = 10^9$ M$_\odot$ (orange). Top  Right: Galaxy bias, same color code as on the left hand side. In the ideal case the theoretical bias is $b = 5$, in the   cases with Poisson noise we incorporate the redshift-dependent bias  $b(z)$. Bottom: $\hat W(k,z)$  (data points) at $z = 14$ (left) and $z = 9$ (right)  without (black) and with (purple and orange) Poisson noise. Curves mark   $W(k,z)$  (solid/dashed lines correspond to the  positive/negative values of $W(k,z)$). }
\label{fig:Pois1} 
\end{figure*}

So far we have assumed an idealized galaxy survey. In this section we consider a realistic galaxy survey with added Poisson fluctuations (see Section \ref{Sec:gals} for details) in the (more realistic) case of the
non-linear velocity. The fitting procedure is the same as described in Section \ref{Sec:Fitting}. The results are shown in Figure  \ref{fig:Pois1} for M$_{\rm min} = 10^8$  M$_\odot$ (purple)  and M$_{\rm min} = 10^9$  M$_\odot$ (orange) together with the ideal case (black) which is shown for comparison. The theoretical expectations are shown with solid curves. Note that the theoretical predictions for the bias  depend on the minimum cutoff halo mass;  while the theoretical curves for the global signal and the window function are independent of M$_{\rm min}$.

The strong bias of bright galaxies makes the extraction of astrophysical parameters much more challenging, especially at high redshifts where these sources are extremely rare (as can be seen from Figure \ref{fig:bn}  which shows the cosmic mean halo number in each cell  as a function of redshift and for several choices of M$_{\rm min}$). This is because  in Eq. \ref{Eq:Rk},
if the bias is  large, the anisotropic terms become relatively smaller and harder to measure. This effect combines
with the larger Poisson noise of the rare galaxies.
Although subject to large errors, the method still can be applied when M$_{\rm min} $  is low, e.g.,  M$_{\rm min} = 10^8$  M$_\odot$. In this case major features of $T_{21}(z)$, $b(z)$ and $W(k,z)$ can still be measured (purple data points in Figure \ref{fig:Pois1}).  For instance, although there is an offset between the theoretical and the measured window function, its shape is recognized correctly and the amplitude is within an order of magnitude of the theoretical value. The discrepancy  is a few tens of percent in both the global signal and the redshift-dependence bias. However, the quality of the fit dramatically degrades  as the cutoff mass is increased. Increasing M$_{\rm min}$ by just an order of magnitude greatly weakens the anisotropy signature and completely impedes parameter estimation, so that none of
the quantities are measured correctly (orange data points in Figure \ref{fig:Pois1}).

\begin{figure*}
\includegraphics[width=3.2in]{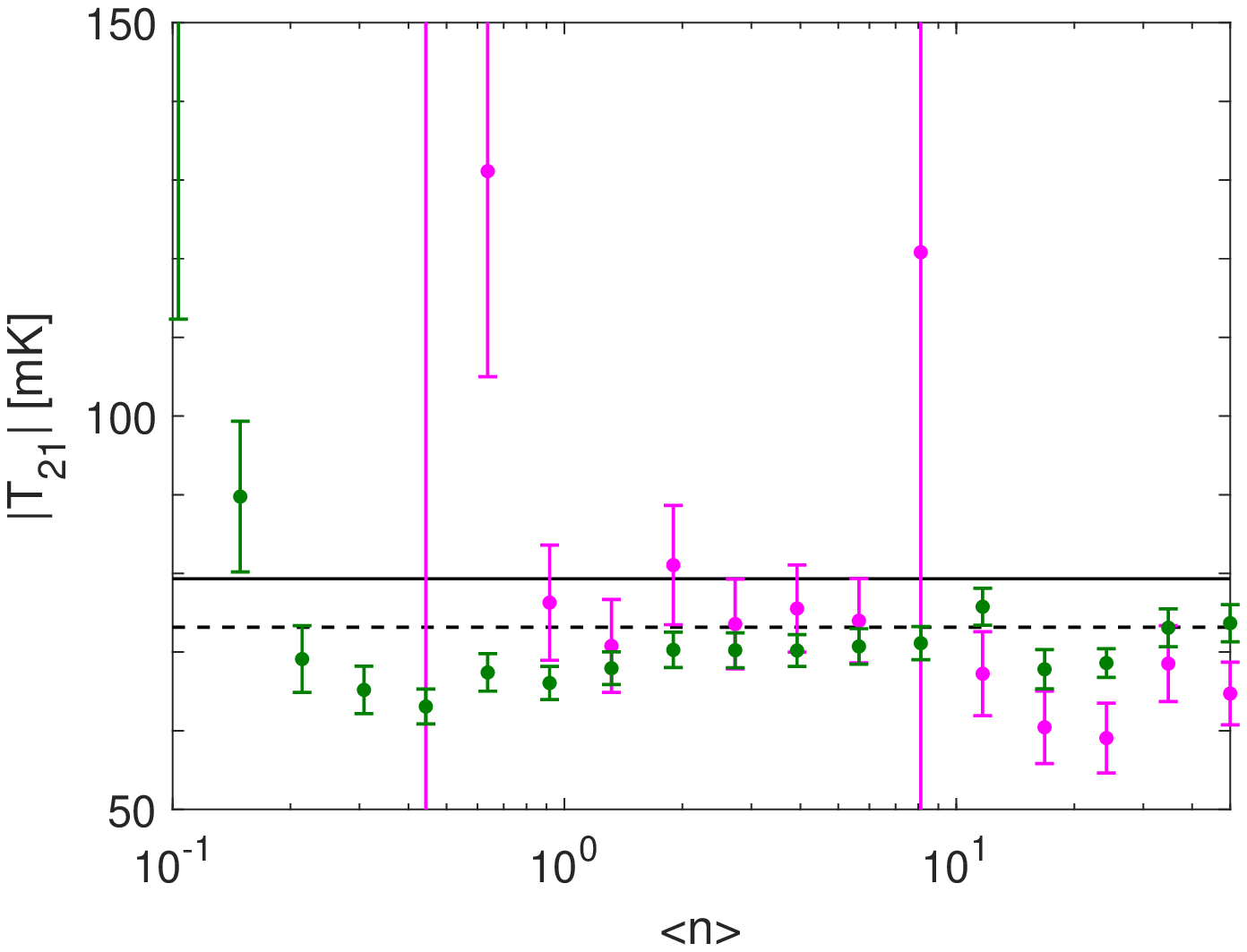}\includegraphics[width=3.2in]{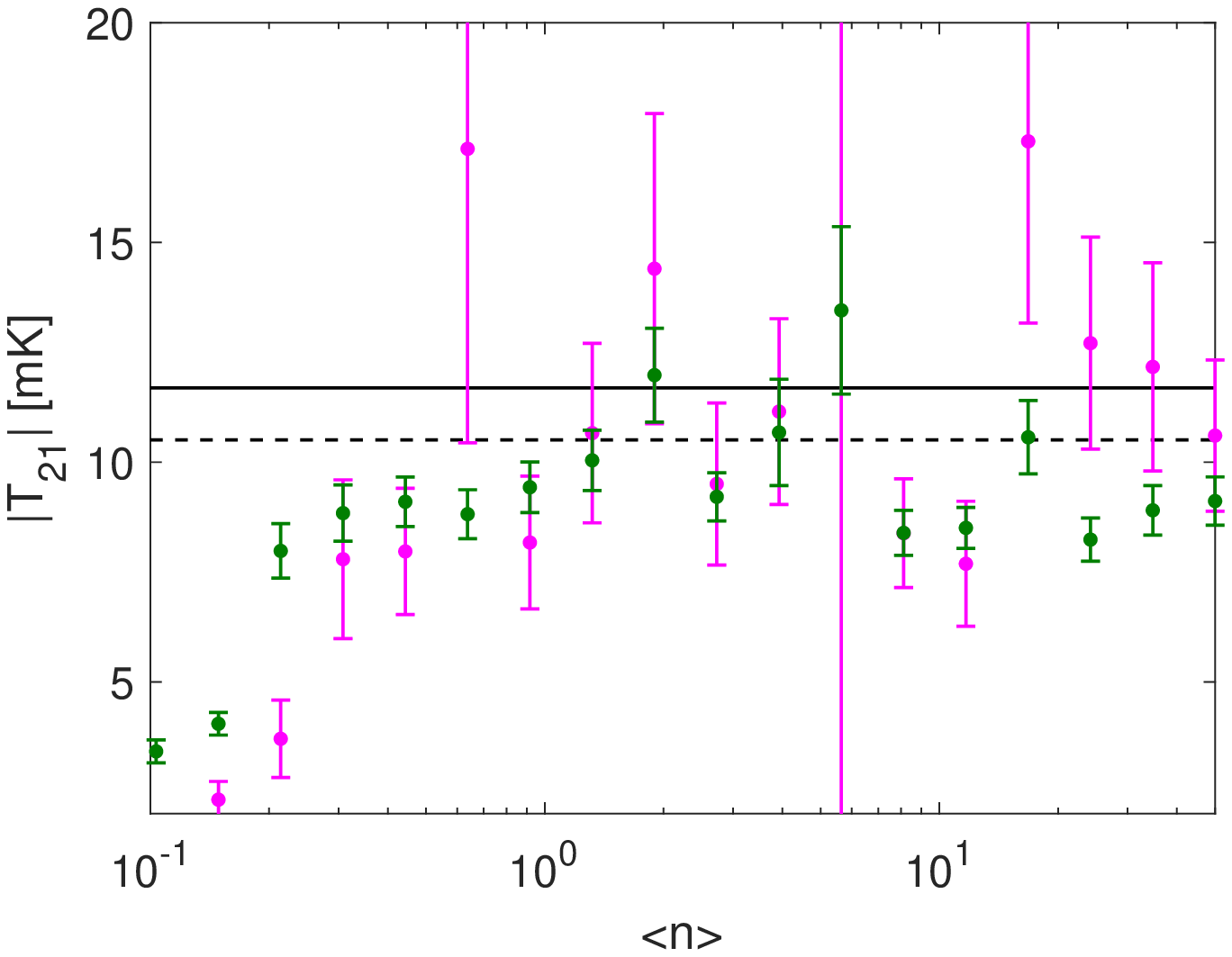}
\caption{Extracting $T_{21}$ with Poisson fluctuations (data points) from galaxy samples with modified values of  $<n>$ (in units of Mpc$^{-3}$)  varied between 0.05 and 50 from a small box (128$^3$, magenta) and a large box (256$^3$, green) at $z = 15$ (left) and at $z = 9$ (right). At each redshift we show the  real $|T_{21}|$ (Case2) which is constant as a function of $<n>$ (horizontal line, solid black). For comparison we also show $|T_{21}|$ extracted from an ideal galaxy sample (horizontal line, dashed black, large box).   }
\label{fig:en} 
\end{figure*}

Next, we test  the effect of the Poisson noise across the halo mass scales (assuming redshift-independent bias $b = 5$). To get a feeling of what kind of galaxy surveys would be useful, we artificially increase $<n>$ and show the absolute value of the reconstructed temperature $T_{21}(z)$ at redshifts 14 (left) and 9 (right)  versus the value of $<n>$ in Figure \ref{fig:en} for  a small box (128$^3$, magenta) and a large box (256$^3$, green). As expected, because of the lower values of the bias at  lower M$_{\rm min}$ (equivalent to larger number of observed galaxies) the fitting procedure works better as we increase $<n>$:  at large  $<n>$  the data points sit close to the value of the global signal extracted using the ideal galaxy survey (horizontal dashed line). For the large box, cosmic variance is small, and Poisson fluctuations
are also small as long as $<n>$ is larger than 0.2 Mpc$^{-3}$. For reference, the values of $<n>$ corresponding to M$_{\rm min}=10^8$  and M$_{\rm min} = 10^9$ at $z = 15$  are $<n>$ =  0.0211 Mpc$^{-3}$ and 0.00023 Mpc$^{-3}$; and $<n>$ = 0.3319 Mpc$^{-3}$ and 0.0158 Mpc$^{-3}$ respectively at $z =9$. 

\begin{figure}
\includegraphics[width=3.2in]{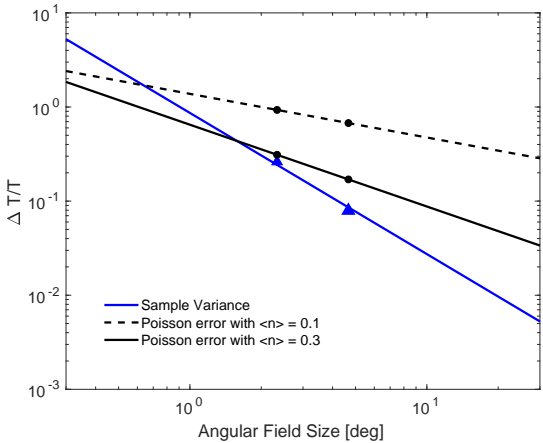}
\caption{Trade-off between sample variance and Poisson noise at $z = 9$. The relative error in the brightness temperature ($\Delta T/T$) is shown 
as a function of survey angular size. The contribution from sample variance (blue solid line)   scales as the survey size to the power $-3/2$, which agrees well with the measured errors (triangles indicate mean values extracted from two small (384 Mpc) and two large (768 Mpc) simulations). Black circles show the Poisson noise extracted from Fig. \ref{fig:en} for $<n> = 0.1$ Mpc$^{-3}$ and $<n> = 0.3$ Mpc$^{-3}$. A linear fit to these data points in $\log_{10}($angular size$)$ vs $\log_{10}({\rm \Delta T/T})$ is shown with dashed and solid black lines respectively.  }
\label{fig:SV} 
\end{figure}

When planning future synergetic missions, one needs to take into account the trade-off between sample variance and Poisson noise. In Figure \ref{fig:SV} we show the relative errors in the brightness temperature from sample variance (blue curve and markers) and Poisson noise (black curves and markers) as a function of survey angular size. As was explained above (and demonstrated in Figure \ref{Fig:CB}), sample variance contributes statistical errors that scale as the angular size to the power of $-3/2$. It is more difficult to estimate the scaling of the Poisson errors based only on the two simulated points (extracted from the small and large simulated boxes) for each $<n>$. To provide rough guidance, here we simply extrapolate the linear fit in $\log_{10}($angular size$)$ vs $\log_{10}({\rm \Delta T/T})$.  We see that small survey sizes are limited by sample variance, while on large angular scales, Poisson noise dominates even for high values of $<n>$, i.e., low bias. However, the dependence of the Poisson error on the survey size steepens with $<n>$ and is expected to be sub-dominant in the case of very low bias.

\section{Conclusions}
\label{Sec:sum}

In this paper we have developed a new method to measure the global spectrum using fluctuations of the 21-cm signal. The method is based on the multi-tracer technique and employs a ratio of two density fields, one of which  is the Fourier transform of a 21-cm map. 
The second map can be of any other tracer probing the same large scale matter density field.   As a proof of concept here we used mock high-redshift galaxy surveys. In the linear regime, the ratio is  anisotropic and  has a well-known angular dependence. We showed that fitting the ratio as a function of the cosine of the angle with respect to the line of sight allows a  measurement of large-scale properties of the two tracers such as the global 21-cm signal, window function of the dominant radiative sources that drive the signal and the galaxy bias. We test the method using both linear and non-linear theory and apply it to an ideal galaxy survey  as well as to a realistic survey which contains only bright strongly biased galaxies.  We use  standard astrophysical scenarios to model the 21-cm signal and do not take the anomalously strong EDGES-Low detection into account. 

The method is reliable when applied to  an ideal galaxy survey and works well  both in the linear and non-linear cases allowing us to measure the global 21-cm signal and the galaxy bias. Reconstruction of the window function is more difficult in the non-linear case although its shape  and sign are recognized correctly. The shape of the window function is directly related to the  spectral energy distribution of the dominant sources that drive the 21-cm signal. In the examined cases the statistical error in the extracted window functions is small enough and allows us to discriminate between hard and soft X-ray sources responsible for cosmic heating. 

We showed that at $z\lesssim 15$ the method can be applied to realistic galaxy samples featuring  deep integration and a large field of view. For example, the reconstruction worked reasonably well when applied to an SKA-size survey area over which all galaxies in halos of $10^8$ M$_\odot$ and above were counted  (corresponding to M$_* \geq  7 \times 10^5$ M$_\odot$). A higher mass threshold implies a sparser and more strongly biased sample of galaxies and reduces the effectiveness of the method. Considering redshifts higher than $z \sim 15$ is not practical when applying the presented method to future galaxy surveys such as the 2000 deg$^2$ Large Area Near Infrared Survey with WFIRST which is expected to detect 1000 galaxies at $z>10$ at the $5\sigma$ limit (and only 100 at the $10\sigma$ limit), deep observations with {\it JWST} which will probe much smaller solid angles and deep surveys with Euclid which will cover  $\sim 40$ deg$^2$ with thousands  of galaxies at $z>7$. 

The method can be extended and applied to other tracers of the underlying large-scale density field. For example,  CO or [CII] intensity mapping of high-redshift galaxies during the EoR  \citep[$z\sim 6-10$,][]{Moradinezhad:2019b, Moradinezhad:2019} is a potentially interesting candidate for such a study. Also, the population of resolved X-ray sources probed by {\it Lynx}, as well as the unresolved cosmic X-ray background, could play the role of the second tracer.  If it is approved, {\it Lynx} will be sensitive to $\sim 10^4$ M$_{\odot}$ accreting black holes at redshift out to  $z\sim 10$. If a wide enough survey with such an instrument turns out to be possible, the multi-tracer technique could be applied to the ratio of the 21-cm fluctuations and a map of X-ray sources at $z\sim 10$. On the other hand, the unresolved cosmic  X-ray background could help to constrain  the global spectrum of the  21-cm signal from cosmic dawn as well as to directly measure the relevant parameters of the population of X-ray binaries that heated up the IGM.

 Our results are timely as more high-redshift probes are coming online, and encourage synergy between the upcoming high-redshift surveys across the entire electromagnetic spectrum. Ours is a proof-of-concept study, and
further investigation is needed. First, exploring larger fields  and stronger signals (in accordance with EDGES-Low) could make the method work better. Second, our non-linearity is only partly realistic, and a full numerical simulation will likely have
more strongly non-linear fields (both 21-cm and galaxies). Finally, in this paper we assumed perfect foreground avoidance/removal, and more tests including realistic foreground treatment are required.

\section{Acknowledgments}

We acknowledge the usage of the Harvard {\it Odyssey} cluster. Part of this project 
was supported by the Royal Society University Research Fellowship of AF. 
This project/publication was made possible for RB through the support
of a grant from the John Templeton Foundation. The opinions expressed
in this publication are those of the authors and do not necessarily
reflect the views of the John Templeton Foundation. RB was also
supported by the ISF-NSFC joint research program (grant No. 2580/17).
This work was begun under a Leverhulme Trust
Visiting Professorship for RB at the University of Oxford.


\appendix
\section{Technical Details of the Fitting Procedure}
\label{App1}
\begin{figure}
\begin{center}
\includegraphics[width=2.8in]{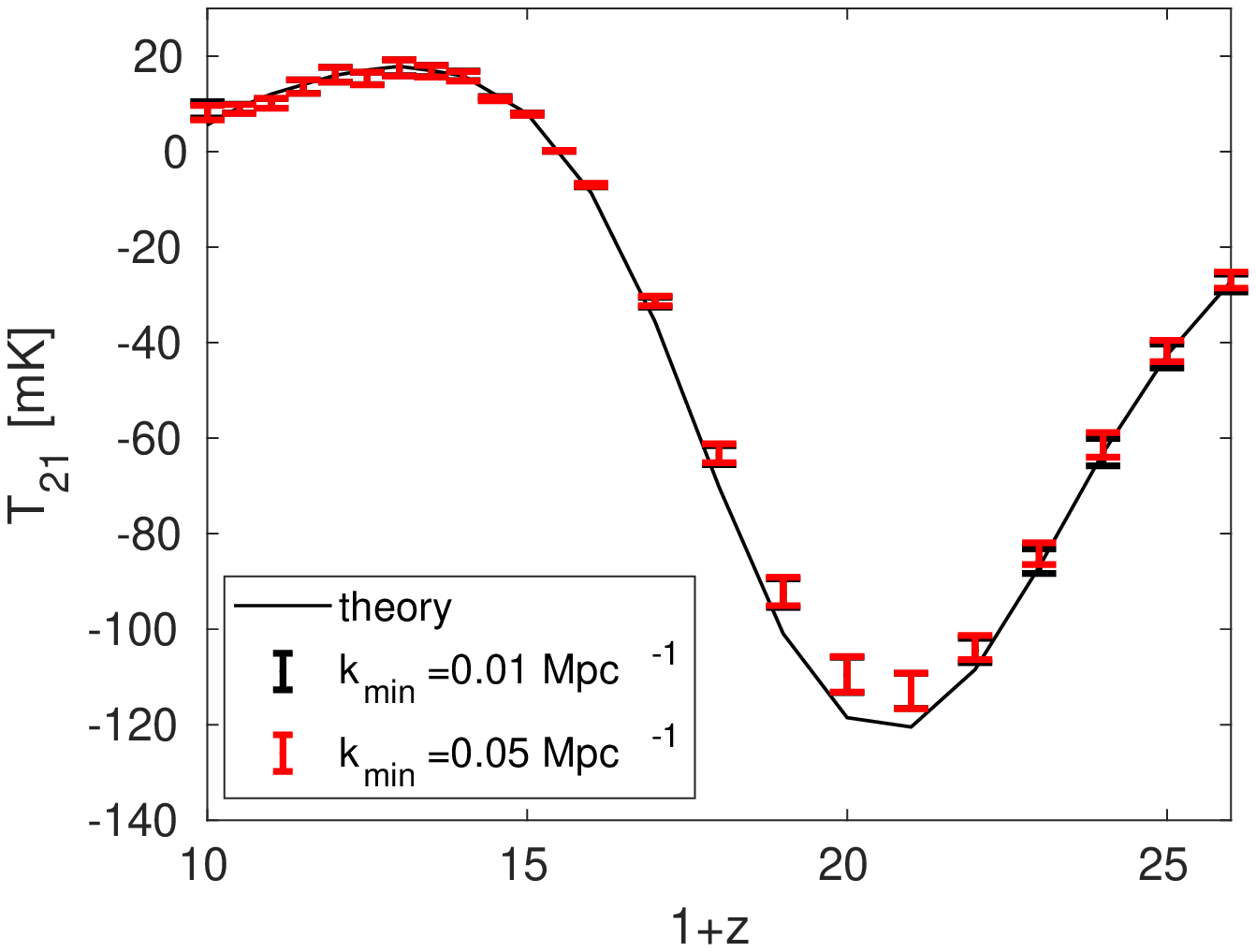}
\includegraphics[width=2.8in]{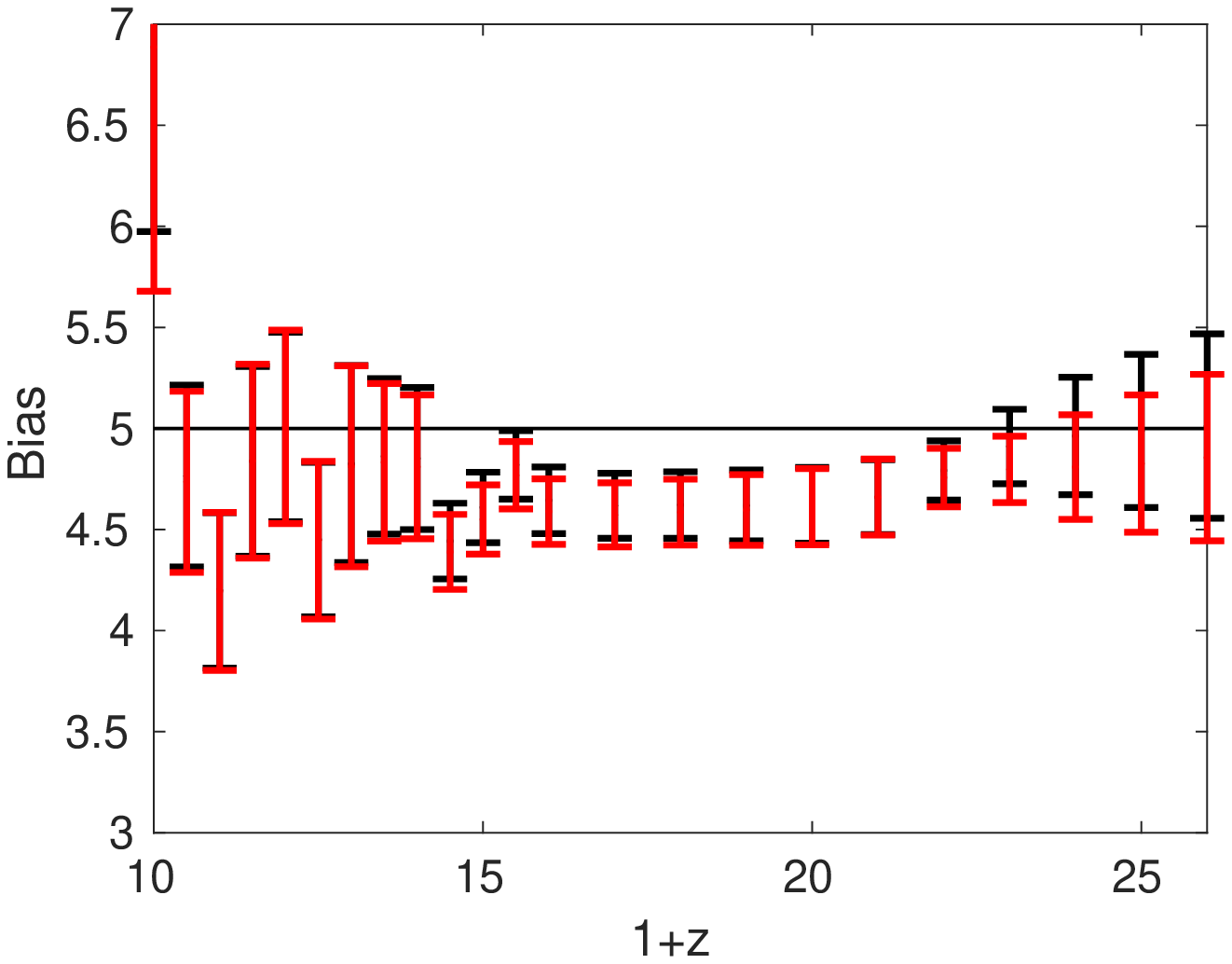}
\includegraphics[width=2.8in]{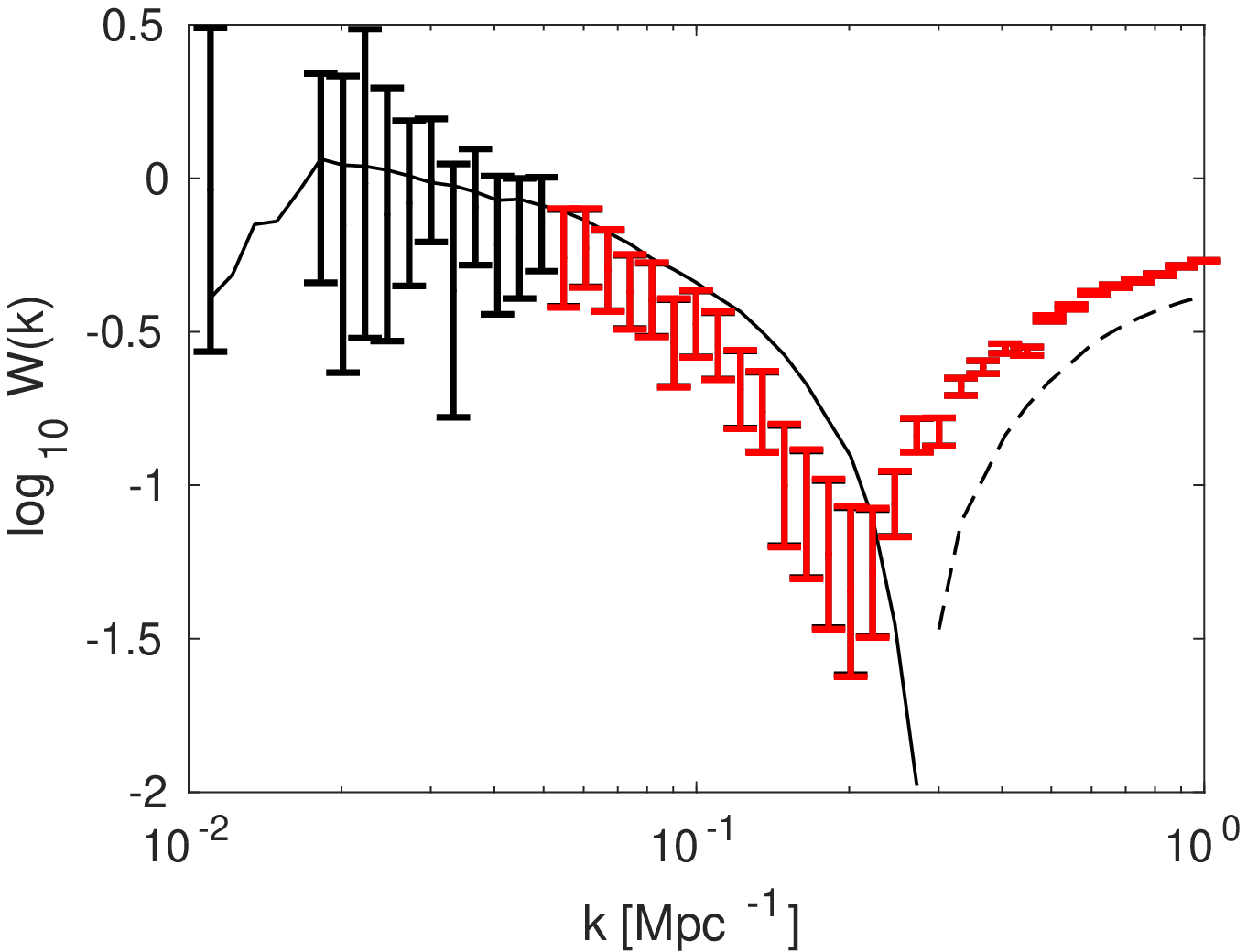}
\caption{Robustness of the multi-tracer method  to variations in the range of wavenumbers. Global 21-cm signal (top), galaxy bias (middle) and window function at redshift $z = 12$ (bottom) are shown for the Case 3 model. Solid/dashed lines show theoretical predictions (dashed is used only for the negative values of $W(k,z)$); data points correspond to  $k = 0.01-1$ Mc$^{-1}$ (black) and 
$k = 0.05-1$ Mc$^{-1}$ (red). }
\label{Fig:app} 
\end{center}
\end{figure}

In the process of fitting, described in Section \ref{Sec:Fitting}, we encountered several technical challenges:
\begin{itemize}
\item It is important to find a  sweet-spot between the number of bins along the $X=\mu^2$ axis, $N_X$, and the number of data points per X-bin to guarantee both the smallest possible statistical errors in the estimated parameters and the good quality of the fit. We tested the convergence of the results for $N_{X}$ between 5 and 150 and obtained the best results for $N_{X} = 50$. Because the number of samples per one $X$-bin is smaller at larger scales (smaller $k$), statistical errors are larger in  $\hat W(k,z)$ at lower values of $k$.
\item We find that fitting the nonlinear function  of X in Eq. \ref{Eq:Fit} required setting physically motivated priors and making an initial guess (within the physically motivated range) for the parameters $B$ and $A_k$  (while the procedure is not sensitive to the initial value of $C$ as long as the value is positive). The fitting procedure is, therefore, performed iteratively in a few steps:
\begin{enumerate}
\item We start by computing the ratio of our simulated fields
 (either in the linear or in the non-linear regime) for a range of $k$ and as a function of $X$ at a given redshift $z$.
\item Fixing  the value of $C$ (we chose to always start with $C=5$), we multiply the ratio by $(C+X)$. For every  $k$ we fit the product with a linear function and find the $y-$intercept (which is our first estimate of $A_k$) and  the slope (which gives $N_k$ estimates for $B$ out of which we use the mean value). These  values of $A_k$ and $B$ are used as initial guesses at the next step. We also use the analytical calculation of $b(z)$ as the initial guess for galaxy bias when fitting the ratio in the case of a realistic galaxy sample.
\item At the next stage a non-linear weighted fit (weights are proportional to the  number of samples per bin) is performed simultaneously for all $k$.  $N_k+2$  parameters are estimated, and the values of $\hat b(z)$ and  $\hat T_{21}(z)$ together with $\sigma_b$ and $\sigma_T$  are derived. At this point we do not save the value of $\hat W(k,z)$. 
\item Using the value of $C = \hat b(z)$ calculated above, we again multiply the ratio by $(C+X)$ and  fit with a linear function to find $\hat W(k,z)$. We estimate the error as $\sigma_W = \sigma_{y-\rm intercept}/\hat T_{21}(z)$. 
\end{enumerate}
At the end of this procedure we have the  best fit values of $\hat b(z)$, $\hat T_{21}(z)$ and $\hat W(k,z)$ together with $1-\sigma$ errors in each parameter.  
\item  We have tested that the fitting procedure is robust to small changes in the range of wavenumbers. An example is shown in Figure \ref{Fig:app}, where  $\hat T_{21}(z)$, $\hat b(z)$  and $\hat W(k,z)$ are the results of fitting in the range $k = 0.01-1$ Mpc$^{-1}$ (black) and 
$k = 0.05-1$ Mpc$^{-1}$ (red). The agreement between the two sets of results is excellent. We find that the reconstructions of both the global signal and the galaxy bias work equally well with and without the low k-modes.   This test also shows that even if the lowest wavenumbers are lost to foreground subtraction, the method will work equally well as long as the higher k-modes are probed with good enough  precision (i.e., we still require large field of view).
\end{itemize}

\end{document}